\documentclass[11pt]{article}
\pdfoutput=1

\usepackage{jheppub} 

\usepackage{slashed,bm}
\usepackage{array,multirow,amsmath,latexsym,bm,graphicx}
\allowdisplaybreaks 
\usepackage[normalem]{ulem}
\usepackage[T1]{fontenc}
\usepackage{multicol}

\usepackage{mathrsfs}

\def \beq{\begin{equation}}
\def \eeq{\end{equation}}
\def \beqa{\begin{eqnarray}}
\def \eeqa{\end{eqnarray}}

\def\msbar{$\overline{\hbox{MS}}$}

\def\sc{{\overline s}}

\def\Ct{{\tilde{C}}}
\def\mCt{{\tilde{\mathcal{C}}}}

\def\bea{\begin{eqnarray}}
\def\eea{\end{eqnarray}}

\subheader{\small   IPPP/20/31}

\title{Higgs decay to fermion pairs at  NLO in SMEFT}

\author{Jonathan~M.~Cullen,}
\author{Benjamin~D.~Pecjak}

\affiliation{Institute for Particle Physics Phenomenology, 
Durham University,  Durham DH1 3LE, UK}

\abstract{The calculation of next-to-leading order (NLO) perturbative corrections at fixed operator dimension in 
Standard Model Effective Field Theory (SMEFT) has been a topic of much recent interest.  
In this paper we obtain the NLO corrections from dimension-6 operators to the 
Higgs boson decays $h\to f\bar{f}$, where the fermions $f \in \{\mu,\tau,c\}$.  This extends previous results
for $h\to b\bar{b}$ to all phenomenologically relevant Higgs boson decays into fermions, and provides the basis
for future precision analyses of these decays within effective field theory.
We point out the benefits of studying ratios of decay rates into different fermions in SMEFT, the most surprising
of which is enhanced sensitivity to anomalous $h\gamma\gamma$ and $hgg$ couplings induced by flavor-universal
SMEFT operators, especially in scenarios where flavor-dependent Wilson coefficients are constrained by 
Minimal Flavor Violation. }

\emailAdd{jonathan.m.cullen@durham.ac.uk}
\emailAdd{ben.pecjak@durham.ac.uk}

\keywords{Higgs physics, Effective Field Theory}

\begin{document} 
\maketitle
\flushbottom

\section{Introduction}

Higgs decay into a fermion anti-fermion pair ($h\to f\bar{f}$) is a benchmark process in the Standard Model (SM).
At present, the LHC has measured Higgs decays into $b$-quarks \cite{Aaboud:2018zhk,Sirunyan:2018kst} and $\tau$-leptons \cite{Sirunyan:2017khh,Aaboud:2018pen}, both with an $\mathcal{O}(10 \%)$ precision,  
and put limits on decays into muons \cite{Aaboud:2017ojs,Sirunyan:2018hbu} and charm quarks \cite{Aaboud:2018fhh,Sirunyan:2019qia}.    At lepton colliders, decays into $b$-quarks,  $c$-quarks, and 
$\tau$-leptons should reach percent-level precision, with somewhat worse performance
for decays into muons \cite{Ruan:2014xxa,Fujii:2015jha,Abramowicz:2016zbo,Ellis:2017kfi,Durieux:2017rsg}.

As measurements become more precise, $h\to f\bar{f}$ decays will allow the measurement of the SM Yukawa  
interactions and place strong constraints on non-standard Higgs couplings.  
Assuming that any potential new physics (NP) contributions come only from particles with masses 
much heavier than the electroweak (EW) scale, Standard Model Effective Theory (SMEFT) offers a model-independent,
systematically improvable quantum-field theoretical framework for analyses of Higgs decays beyond the SM.  One possible
improvement, which has received much interest of late,  is to include next-to-leading order (NLO) perturbative corrections 
at a fixed dimension in the operator expansion \cite{Pruna:2014asa,Hartmann:2015oia,Hartmann:2015aia,Hartmann:2016pil,Ghezzi:2015vva,Dawson:2018pyl,Dawson:2018liq,Dawson:2018jlg,Dawson:2019clf,Deutschmann:2017qum,Grazzini:2018eyk,Boughezal:2019xpp,Dedes:2019bew,Vryonidou:2018eyv,Baglio:2017bfe,Dawson:2018dxp,Degrande:2016dqg,Zhang:2014rja,Zhang:2013xya,Zhang:2016omx,Degrande:2018fog,Bylund:2016phk,Maltoni:2016yxb,Grober:2015cwa,Neumann:2019kvk,deFlorian:2017qfk,Crivellin:2013hpa,Dedes:2018seb}.

In \cite{Cullen:2019nnr}, we built upon the results of \cite{Gauld:2015lmb, Gauld:2016kuu} to calculate
the full set of NLO corrections from dimension-6 SMEFT operators to the decay 
rate $h\to b\bar{b}$.  On the phenomenological side, those results provide a baseline for future precision studies 
of that decay mode using effective field theory.  From the perspective of SMEFT development,
that paper laid out  a hybrid renormalization scheme appropriate for combining EW and QCD 
corrections.  The main idea, outlined in Section~\ref{sec:set-up}, was to use the ``FJ tadpole scheme'' \cite{Fleischer:1980ub} within SMEFT to define gauge-independent \msbar~renormalized light fermion masses and QED/QCD gauge couplings, while all other masses are renormalized on-shell.  
Expressing the decay rate in terms of the corresponding \msbar~parameters defined 
in a five-flavor version of QCD$\times$QED requires one to add on to the decay rate virtual corrections involving
the top quark and heavy electroweak bosons characteristic of on-shell renormalization  
through terms involving perturbatively calculable decoupling constants.  
The benefit of this procedure is that a set of large logarithms of the ratio $m_f/m_H$ appearing in 
QCD$\times$QED corrections is resummed into the definitions of the \msbar~parameters, 
while spuriously large tadpole corrections from top-quark loops cancel from the decay rate, as in a purely on-shell scheme.

The goal of the present work is to apply that framework to obtain the NLO dimension-6 SMEFT corrections to 
Higgs decays into charm quarks, $\tau$ leptons, and muons, thus extending the results of  \cite{Cullen:2019nnr} to cover 
the full spectrum of phenomenologically relevant $h\to f\bar{f}$ decays.
While these cases are conceptually similar to $h\to b\bar{b}$, the explicit 
Feynman diagrams and dimension-6 Wilson coefficient entering the calculation differ, 
necessitating us to calculate from scratch the large majority  of EW corrections.  After setting some notation for mass-basis Wilson coefficients, also in Minimal Flavor Violation (MFV), in Section~\ref{sec:WilsonCoefficientsMFV}, we give a small subset of the analytical results in Section~\ref{sec:AnalyticalResults}, accompanied by illustrative numerical studies in Section~\ref{sec:NumericalResults}.  The complete analytical results are rather  lengthy and included in electronic form with the electronic submission of this article.  In addition to our analysis of individual  decay channels, we emphasize in Section~\ref{sec:Ratios} the advantages of studying ratios of decay modes in SMEFT.  In particular, since a large number of Wilson coefficients drop out of ratios, particularly if MFV is  imposed, ratios of decay rates may allow for stronger constraints on Wilson coefficients than decay rates alone.  We summarize our findings in Section~\ref{sec:Conclusions}, and in Appendix~\ref{sec:DecouplingConstants} we list results for the decoupling constants used in the renormalization
procedure.

\section{Calculational set-up}
\label{sec:set-up}
In this section we outline the procedure used to obtain the NLO corrections to the decay rates $h\to f\bar{f}$, with
$h$ the Higgs boson and $f$ a fermion,  in dimension-6 SMEFT.  The results depend on a set of input parameters 
and the renormalization scheme in which they are defined.  A main outcome of the NLO SMEFT calculation of
 $h\to b\bar{b}$  decays in~\cite{Cullen:2019nnr} was the development of a hybrid renormalization scheme that minimizes 
 the impact of potentially large higher-order corrections that appear when naively combining QCD and EW
 corrections.  It uses as input the following parameters:
\begin{equation}
\label{eq:InputPars}
m_{EW},  \space \overline{m}_f, \space \overline\alpha, \space \alpha_s,  \space C_i,  \space  V_{ij}\,    .
\end{equation}
Some comments on the input parameters and notation are in order:
\begin{itemize}
\item The massive electroweak bosons and top quark are considered as heavy particles with masses 
denoted collectively by $m_{EW}\in \{m_W, \, m_Z, \, m_H, \,m_t\}$.  These are renormalized in the on-shell scheme.
\item The light fermion masses  $\overline{m}_f\equiv \overline{m}_f(\mu)$, with $f$ any fermion other than the top quark,
are renormalized in the \msbar~scheme in a 5-flavor version of QED$\times$QCD, where the top quark is 
integrated out. The same is true of the electromagnetic fine-structure and strong coupling constants,
 $\overline{\alpha}=\overline{e}^2/(4\pi)\equiv \overline{\alpha}(\mu)  $ and $\alpha_s=g_s^2/(4\pi)\equiv \alpha_s(\mu)$.
 In all cases $\overline{m}_f\ll m_{EW}$, and to simplify results we work to first non-vanishing order in 
 this ``small-mass limit''.\footnote{Decay into first-generation fermions is not considered and we set their
 masses to zero throughout the paper.}  The decay rates are finite in this limit, with the exception of double logarithms
 in the ratio $\overline{m}_f/m_H$ appearing in NLO SMEFT corrections proportional to effective $h\gamma\gamma$ and $hgg$ vertices, see Section~\ref{sec:QEDxQCD}. We have checked that corrections to the small-mass limit
 are typically at or below the percent level, depending on the particular Wilson coefficient.  
 
 In \cite{Cullen:2019nnr} the alternative notation $\overline{m}_f^{\ell}$ and $\overline{\alpha}^\ell$ was used
 for \msbar~parameters renormalized in 5-flavor  QED$\times$QCD, in order to make explicit the difference between the  corresponding parameters in the full SM,  where heavy particles also contribute to the running. 
We omit the superscript $\ell$ in the present work, with the understanding that the running 
 of \msbar~parameters is driven exclusively by light particles.  Effects of heavy particles are integrated
 out and added on to the decay rate perturbatively through terms involving decoupling constants as discussed in 
 Appendix~\ref{sec:DecouplingConstants}. 
 
 \item The $C_i \equiv C_i(\mu)$ are Wilson coefficients of the dimension-6 SMEFT operators $Q_i \equiv Q_i(\mu)$, for which we use the Warsaw basis~\cite{Grzadkowski:2010es} in Table~\ref{op59}.  The Wilson coefficients are renormalized in the \msbar~scheme, with all SM particles contributing to the running.  We use the notation where 
 the dimension-6 piece of the SMEFT Lagrangian is ${\cal L}^{(6)}=\sum C_i Q_i$, so the $C_i$ have mass dimension minus 2.
\item As standard in EW calculations,  we use the numerical approximation where the CKM matrix $V_{ij}=\delta_{ij}$. 
\end{itemize}

The decay rate up to NLO in perturbation theory is written as 
\begin{align}
\label{eq:PertExpansion}
\Gamma(h\to f\bar{f})\equiv \Gamma_f & = \Gamma_f^{(0)}+\Gamma_f^{(1)}  \, , 
\end{align}
where the superscripts (0) and (1) refer to the  LO and NLO contribution in perturbation theory.
We split these contributions into an SM part and a piece containing exactly one 
Wilson coefficient of a dimension-6 operator as
\begin{align}
\label{eq:PertExpansionSMEFT}
\Gamma_f^{(0)}  & = \Gamma_f^{(4,0)} +\Gamma_f^{(6,0)}  \, ,  \nonumber \\
\Gamma_f^{(1)} & =  \Gamma_f^{(4,1)} +   \Gamma_f^{(6,1)} \,,
\end{align}
so that the double superscripts $(i,j)$ refer to the dimension-$i$ contribution at $j$-th order in perturbation theory.   
It is useful to subdivide the NLO corrections into two pieces as 
\begin{align}
\label{eq:PertExpansion2}
\Gamma_f^{(i,1)}& = \Gamma^{(i,1)}_{f, (g,\gamma)}+  \Gamma_{f,\rm weak}^{(i,1)}  \, .
\end{align}
The $\Gamma^{(i,1)}_{f,(g,\gamma)}$  contain 1-loop diagrams involving at least one photon or gluon propagator, in addition to
contributions from the three-body processes $h\to f\bar{f}(g,\gamma)$, whereas  $\Gamma_{f,{\rm weak}}^{(i,1)}$ contains
the remaining NLO corrections. 

The procedure for obtaining $\Gamma_f^{(1)}$ in the hybrid renormalization scheme described above was set forth in 
\cite{Cullen:2019nnr}, and applied to the case of $h\to b\bar{b}$.\footnote{The SM result $\Gamma_f^{(4,1)}$ was first 
calculated in on-shell scheme for all parameters in \cite{Kniehl:1991ze}.}  It included a careful treatment of tadpoles in  
the  FJ-tadpole scheme, and the calculation of decoupling constants relating \msbar-renormalized quantities in 
5-flavor QCD$\times$QED to the corresponding quantities in the full SM.  
The benefit of this renormalization scheme is threefold.  First,  it treats contributions from light-particle loops, which are sensitive both to $m_{EW}$ and $\overline{m}_f$, in the \msbar~scheme, thus  resumming a set of corrections 
of the form $\ln \overline{m}_f/m_H$ that appear in $\Gamma^{(i,1)}_{f,(g,\gamma)}$ 
through the use of the  running mass $\overline{m}_f$.  Second, contributions from heavy-particle loops depending on  
$m_{EW}$ appear in $\Gamma_{f,{\rm weak}}^{(i,1)}$ through  decoupling constants for $\overline{m}_f$ and $\overline{\alpha}$; these pieces are effectively calculated in the on-shell scheme, where tadpoles cancel between different terms in the decay rate, so enhanced EW corrections scaling as $m_t^4/m_H^2 \overline{v}^2$ (where $\overline{v}$ is defined in Eq.~\ref{eq:SMVar} below) due to these tadpoles are absent.  Finally,  the running of the parameters $\overline{m}_f$ and $\overline\alpha$ does not depend on the SMEFT Wilson coefficients $C_i$ to leading order in $\overline{m}_f/m_{EW}$, so these are easily related to \msbar-renormalized parameters extracted from low-energy experiments and LEP.

There are no conceptual complications in extending the methods for $h\to b\bar{b}$ to other cases such as $h\to c\bar{c}$ or $h\to \tau\bar{\tau}$.  Indeed, the NLO results for $\Gamma^{(i,1)}_{f,(g,\gamma)}$, as well as the contributions of four-fermion
operators to $\Gamma^{(i,1)}_{f,\rm weak}$ can be obtained straightforwardly from \cite{Cullen:2019nnr,Gauld:2015lmb,Gauld:2016kuu}.  We give these partial NLO results in analytic form in Section~\ref{sec:AnalyticalResults}.  However,  
the same is not true of $\Gamma^{(i,1)}_{f,\rm weak}$.  These weak corrections receive flavor-dependent contributions from a large set of one-loop diagrams entering mass renormalization, the decoupling constants, and the bare one-loop matrix elements, which we must calculate from scratch. To this end, we have altered the in-house code developed to automate the one-loop $h\to b\bar{b}$ calculation \cite{Cullen:2019nnr}.  It implements the SMEFT Lagrangian in the mass basis, including ghosts, into \texttt{FeynRules}~\cite{Alloul:2013bka}, and then uses  \texttt{FeynArts}~\cite{Hahn:2000kx} and  \texttt{FormCalc}~\cite{Hahn:1998yk} to generate and calculate the one-loop diagrams.  \texttt{Package-X}~\cite{Patel:2015tea} has also been used for analytic cross-checks of certain loop integrals.

The full NLO results are rather lengthy, and are included in computer files in the arXiv 
submission of this work.  They involve contributions from 40 different dimension-6 operators for $h \to \tau \bar{\tau}$, 40
for $h\to \mu\bar{\mu}$, and 47  for $h \to c\bar{c}$. Electronic results for the decoupling constants are given as well, although these are compact enough to list in Appendix~\ref{sec:DecouplingConstants}.  
We have performed the standard consistency checks on our results. In particular, we calculate the  weak corrections in both Feynman and unitary gauge,  check the cancellation of UV and IR poles in the dimensional regulator, and make sure that the decay rate is independent of unphysical renormalization scales up to  NLO.

\section{Wilson coefficients and MFV}
\label{sec:WilsonCoefficientsMFV}

The dimension-6 SMEFT operators in Table~\ref{op59} are defined in the weak basis, but the physical decay rates 
and on-shell renormalization conditions are defined in the mass basis. Moreover, with the exception of the top quark,
Higgs couplings to fermions in the SM are suppressed by small and hierarchical Yukawa couplings, a feature not inherited
by generic SMEFT interactions.  In order to avoid pushing the UV scale of the effective theory to values far above the 
TeV scale to avoid flavor constraints, one often considers the SMEFT Wilson coefficients to be constrained by MFV, 
and we ourselves will study this scenario.  The purpose of this  section is to briefly outline both of these issues, 
and to set up the notation required to implement them in our  results.

We first explain our treatment of Wilson coefficients in the mass basis.  
As an example, consider the term in the Lagrangian involving  the operator $Q_{uH}$ 
defined in the weak basis in Table~\ref{op59}.  The explicit form of this term, which we denote by
${\cal L}_{uH}$, is
\begin{align}
{\cal L}_{uH} =C^w_{\substack{uH \\ rs}}  (H^\dagger H)(\bar{q}^w_r u^w_s \widetilde{H} ) \, .
\end{align}
The superscript $w$ indicates that the Wilson coefficient and fermion fields are defined in the weak basis, and 
the subscripts $r,s$ are flavor indices.  The mass-basis fermion fields are related to  the weak-basis fields
via the unitary transformations
\begin{align}
u^w_R &= U_{u_R} u^m_R \, ,\nonumber \\
q^w_L & = 
U_{u_L}
\begin{pmatrix}
u^m_L  \\
V d^m_L
\end{pmatrix}    \, .
\end{align}
Using that the CKM matrix $V_{ij}\equiv (U_{u_L}^\dagger U_{d_L})_{ij}\approx \delta_{ij}$, 
the Lagrangian contribution can be written as 
\begin{align}
{\cal L}_{uH} =C^m_{\substack{uH \\ rs}}  (H^\dagger H)(\bar{q}^m_r u^m_s \widetilde{H} ) \, ,
\end{align}
where $q^m =(u_L^m,d_L^m)^T$ is a doublet of mass-basis fields and the mass-basis Wilson coefficient is defined as 
\begin{align}
\label{eq:Cm}
C^m_{\substack{uH \\ rs}} \equiv [U^{-1}_{u_L}]_{ri} C^w_{\substack{uH \\ ij}} [U_{u_R}]_{js} \, . 
\end{align}
This pattern holds in general: the SMEFT Lagrangian in terms of mass-basis fermion fields and Wilson coefficients is obtained
by interpreting the fermion fields in the list of operators Table~\ref{op59} to be in the mass basis, and multiplying it with 
a corresponding mass-basis Wilson coefficient, which can be related to the weak-basis one through rotations such as
Eq.~(\ref{eq:Cm}).\footnote{The definition of mass-basis Wilson coefficients beyond the
 approximation $V_{ij}\approx \delta_{ij}$ is not unique.  For one possible choice see~\cite{Dedes:2017zog}.}  

In quoting analytic and numerical results in the coming sections, we always work with mass-basis quantities, and for simplicity 
drop the superscripts $m$ on the fields and Wilson coefficients. Moreover, within our approximations,
$h\to f\bar{f}$ decay is sensitive to a group of generation-diagonal operators involving right-handed
fermion fields.  For these, we use the shorthand notation where e.g. $C_{c H} \equiv C_{\substack{uH \\ 22}}$,  
thus allowing us to suppress flavor indices.  In fact, the only Wilson coefficients appearing in our calculation 
which require explicit flavor indices are the class-7 quantities $C^{(1,3)}_{Hl}$ and $C^{(1,3)}_{Hq}$, in addition to the coefficients of the class-8 four-fermion
operators.

We next turn to our implementation of MFV.   A pedagogical discussion of MFV can be found in, for instance, 
\cite{Grinstein:2015nya}.  Its implementation in SMEFT was discussed in detail in \cite{Alonso:2013hga}. 
In short, imposing MFV in SMEFT constrains the flavor indices of the Wilson coefficients to be carried by certain combinations
of Yukawa matrices.   Upon rotation to the mass basis, Yukawa couplings are converted to powers of 
the fermion masses, which for light fermions $f$ can be expanded in the small-mass limit $\overline{m}_f \ll m_{EW}$.  

As an explicit example we consider the class-5 Wilson coefficients, starting with $C_{uH}$. 
MFV implies that the weak-basis coefficient takes the form \cite{Alonso:2013hga}
\begin{align}
C^w_{\substack{uH \\ rs}}= \left[Y_u G_{uH}\left(Y_d Y_d^\dagger,Y_u Y_u^\dagger \right)\right]_{rs} \,,
\end{align}
where the function $G_{uH}$ is regular in the limit that its arguments go to zero, but  otherwise arbitrary.
In the approximation where the CKM matrix is the unit matrix,  the MFV scaling for the mass basis 
coefficient is obtained by making the replacement $Y_u \to M_u$, where the mass-basis Yukawa 
matrices for $f\in \{e,u,d\}$ are given by 
 \begin{align}
[M_f]_{ij} &= \sqrt{2}\frac{[m_f]_{ij}}{\overline{v}} \, , \nonumber \\ 
[m_e]&=\text{diag}(0,\overline{m}_\mu,\overline{m}_\tau) \, , \nonumber \\
[m_u]&=\text{diag}(0,\overline{m}_c,m_t) \, , \nonumber \\
[m_d]&=\text{diag}(0,\overline{m}_s,\overline{m}_b) \, ,
\end{align}
with $\overline{v}$ defined in Eq.~(\ref{eq:SMVar}).  
The mass-basis Yukawas are diagonal  matrices and their elements vanish in the small-mass limit, 
with the exception of $\left[M_u\right]_{33}$ which is proportional to the top-quark mass and thus 
order one in that limit.  Therefore, to leading order in the small-mass limit, we can write
\begin{align}
\label{eq:TaylorExpansion}
\left[G_{uH}\left(Y_d Y_d^\dagger,Y_u Y_u^\dagger \right)\right]_{ks} = 
\delta_{ks}G_{uH}(0,0) + \delta_{k3}\delta_{s3}
\sum_{k=1}^\infty \frac{y_t^{2k}}{k!} \left(\frac{d^k}{(dy_t^2)^k}G_{uH}(0,y_t^2)\bigg|_{y_t\to 0}\right) \, ,
\end{align}
where $y_t^2 = 2m_t^2/\overline{v}^2$.
It follows that the expansion of the mass-basis coefficient $C_{uH}$  in the small-mass limit
within MFV  is given by   
\begin{align}
\label{eq:CuHMFV}
C_{\substack{uH \\ rs}}\approx \left[M_u\right]_{rk} \left[  \mathcal{C}^1_{\substack{uH \\ ks}} 
+{\cal O} \left(\frac{m^2}{\bar{v}^2}\right)\right]\,  ,
\end{align}
where $m$ is any light-fermion mass and the explicit expression for $\mathcal{C}^1_{uH}$ can be 
read off by matching with Eq.~(\ref{eq:TaylorExpansion}). Here and below the 
superscript $j$ on the calligraphic Wilson coefficients $\mathcal{C}^j_i$ indicates that they multiply
$j$ explicit powers of mass-basis Yukawa matrices.  Note that the object $\mathcal{C}^1_{\substack{uH \\ ks}}$
is flavor-diagonal, but non-universal in the sense that 
$\mathcal{C}^1_{\substack{uH \\ 11}}=\mathcal{C}^1_{\substack{uH \\ 22}}\neq \mathcal{C}^1_{\substack{uH \\ 33}}$.
Similar statements hold for the MFV version of $C_{dH}$ in the small-mass limit, which can be obtained
from the $C_{uH}$ result by the replacement $u\to d$.  

For the corresponding leptonic operator the MFV expression is
\begin{align}
C^w_{\substack{eH \\ rs}}= \left[Y_e G_{eH}\left(Y_eY_e^\dagger \right)\right]_{rs} \,.
\end{align}
All elements of the mass-basis Yukawa coupling $M_e$ vanish in the small-mass limit,  so
the mass-basis coefficient is given by
\begin{align}
C_{\substack{eH \\ rs}}\approx \left[M_e\right]_{rs} \left[\mathcal{C}^1_{eH}
+{\cal O} \left(\frac{m^2}{\bar{v}^2}\right)\right] \, ,
\end{align}
where  $\mathcal{C}^1_{eH}=G_{eH}(0)$ carries no flavor indices and is thus universal,
 in contrast to the quark cases.

It is a straightforward exercise to obtain analogous results for the other Wilson coefficients in MFV in the small-mass
limit.   For our analysis in Section~\ref{sec:Ratios}, the important point is whether, after factoring out $j$ overall Yukawa factors,
the calligraphic Wilson coefficients $\mathcal{C}_i^j$ are flavor universal (as in the case of  
$i=eH$), or flavor non-universal due to contributions from top-quark Yukawas (as in the case of  
$i=dH,uH$).  The flavor-universal cases  used in this paper are
\begin{align}
C_{\substack{eH \\ pr}} &\approx [M_e]_{pr} \mathcal{C}^1_{eH} \, ,   \quad & C_{\substack{Hd \\ pr}} &\approx \delta_{pr} \mathcal{C}^0_{Hd} \,  , \nonumber \\
C_{\substack{eF \\ pr}} &\approx [M_e]_{pr} \mathcal{C}^1_{eF} \, , \quad & C_{\substack{He \\ pr}} &\approx \delta_{pr}   \mathcal{C}^0_{He}  \,   , \nonumber \\
C^{(1,3)}_{\substack{H\ell \\ pr}} &\approx \delta_{pr} \mathcal{C}^{(1,3),0}_{H\ell} \, , \quad & C_{\substack{le \\ prst}} &\approx   \delta_{pr} \delta_{st} \mathcal{C}^0_{le}  \,  ,
\end{align}
where $F$ is any gauge field appearing in the class-6 operators.  The flavor non-universal cases are
\begin{align}
C_{\substack{uH \\ pr}} &\approx [M_u]_{pr} \mathcal{C}^1_{\substack{uH \\ pr}} \, , \quad & C_{\substack{Hud \\ pr}} &\approx [M_u]_{pr} [M_d]_{pr} \mathcal{C}^2_{\substack{Hud \\ pr}}  \, , \nonumber \\
C_{\substack{dH \\ pr}} &\approx [M_d]_{pr} \mathcal{C}^1_{\substack{dH \\ pr}} \, , \quad & C^{(1,8)}_{\substack{qu \\ prst}} &\approx \delta_{pr}\delta_{st}  \mathcal{C}^{(1,8),0}_{\substack{qu \\ prst}} \, , \nonumber \\
C_{\substack{uF \\ pr}} &\approx [M_u]_{pr} \mathcal{C}^1_{\substack{uF \\ pr}} \, , \quad & C^{(1,8)}_{\substack{qd \\ prst}} &\approx \delta_{st}\delta_{pr} \mathcal{C}^{(1,8),0}_{\substack{qd \\ pr}} \, , \nonumber \\
C_{\substack{dF \\ pr}} &\approx [M_d]_{pr} \mathcal{C}^1_{\substack{dF \\ pr}} \, , \quad & C_{\substack{ledq \\ prst}} &\approx [M_e]_{pr} [M_d]_{st} \mathcal{C}^2_{\substack{ledq \\ st}} \, ,  \nonumber \\
C^{(1,3)}_{\substack{Hq \\ pr}} &\approx  \delta_{pr} \mathcal{C}^{(1,3),0}_{\substack{Hq \\ pr}} \, , \quad & C^{(1,8)}_{\substack{quqd \\ prst}} &\approx [M_u]_{pr} [M_d]_{st} \mathcal{C}^{(1,8),2}_{\substack{quqd \\ prst}} \, , \nonumber \\
C_{\substack{Hu \\ pr}} &\approx \delta_{pr} \mathcal{C}^0_{\substack{Hu \\ pr}} \, , \quad & C^{(1,3)}_{\substack{lequ \\ prst}} &\approx [M_e]_{pr} [M_u]_{st} \mathcal{C}^{(1,3),2}_{\substack{lequ \\ st}} \, ,
\end{align}
where there is no implied summation on repeated indices on the right-hand side of the approximations; this notation makes
clear that the calligraphic coefficients are flavor diagonal in the pairs of indices $pr$ and $st$.
We note that while the Wilson coefficients $C^{(1,8)}_{qd}$, $C_{ledq}$ and $C^{(1,3)}_{lequ}$ carry four flavor indices, their corresponding small-mass MFV expansion functions, $\mathcal{C}_i$, are a function of (and therefore only carry) two flavor indices.

It is worth mentioning that all the SMEFT coefficients can also depend on Yukawas through functions of
flavor invariants such as 
\begin{align}
{\rm Tr}\left(Y_e Y_e^\dagger\right),  \, {\rm Tr}\left(Y_d Y_d^\dagger\right), \,  {\rm Tr}\left(Y_u Y_u^\dagger\right) \,.
\end{align}
In the small-mass limit these are either constants, or functions of $y_t^2$.  They can thus be absorbed into the definitions
of the $\mathcal{C}_i^j$ above.  For consistency of notation, we make explicit that Wilson coefficients in classes 1-4 depend on the above invariants and should also be expanded in the small-mass limit. Since those Wilson coefficients 
carry no flavor indices, this amounts to a change of notation $C_{i}\to \mathcal{C}^0_i$, 
where the superscript indicates that the small-mass limit has been taken in the flavor invariants on which the coefficients
can depend.

\section{Analytical results}
\label{sec:AnalyticalResults}
In this section we provide a subset of analytic results for the decay rates $h\to f\bar{f}$.  This also 
allows us to fix some notation used in the rest of the paper.
For instance, in order to express results in terms of the input parameters Eq.~(\ref{eq:InputPars}), we
introduce the derived quantities
\begin{align}
\label{eq:SMVar}
\overline{v} \equiv \frac{2 M_W \hat{s}_w}{\overline{e}}, \qquad \hat{c}_w^2 \equiv \frac{M_W^2}{M_Z^2}, \qquad 
\hat{s}_w^2 \equiv 1-\hat{c}_w^2 \, .
\end{align}
We can then write the LO dimension-4 and dimension-6  contributions to the $h\to f\bar{f}$ decay rates defined in Eq.~(\ref{eq:PertExpansionSMEFT}) as
\begin{align}
\label{eq:LOgam}
 \Gamma_f^{(4,0)}& = \frac{N_c^f m_H }{8 \pi } \frac{\overline{m}_f^2}{\overline{v}^2}\, , \nonumber \\
\Gamma_f^{(6,0)}& = 2 \Gamma_f^{(4,0)}
\left[C_{H\Box}-\frac{C_{HD}}{4}\left(1- \frac{\hat{c}_w^2}{\hat{s}_w^2} \right)
+\frac{\hat{c}_w}{\hat{s}_w} C_{HWB}  
-\frac{\overline{v}}{\overline{m}_f} \frac{C_{fH}}{\sqrt{2}}\right]\overline{v}^2 \, ,
\end{align}
where  $N_c^{f}=1$ if $f$ is a lepton and $N_c^f=3$ if $f$ is a quark. 
To NLO in dimension-6 SMEFT, the decay  rates depend only on the real parts of the dimension-6 Wilson coefficients.  
We leave this implicit, such that $C_{i}\equiv {\rm Re}  (C_{i})$ for any Wilson coefficient $C_i$ that appears in the 
decay rates, not only above but in the rest of the paper.  Furthermore, we work with the mass-basis Wilson coefficients
discussed in Section~\ref{sec:WilsonCoefficientsMFV}, such that $C_{fH}$ multiplies an operator that alters the $hff$ vertex in that basis.

\subsection{QED$\times$QCD corrections}
\label{sec:QEDxQCD}
The QED$\times$QCD corrections may be deduced from the results in \cite{Cullen:2019nnr}.  The results are
 \begin{align}
\label{eq:QEDQCDGam61}
 \Gamma_{f,(g,\gamma)}^{(4,1)} & = \Gamma_f^{(4,0)}\left(\frac{\delta_{f,q} C_F \alpha_s+ Q_f^2 \alpha}{\pi}\right) \left[\frac{17}{4}+\frac{3}{2}\ln\left(\frac{\mu^2}{m_H^2}\right)\right] \, , \nonumber \\
\Gamma_{f,(g,\gamma)}^{(6,1)}& = 
\Gamma_f^{(6,0)} \frac{\Gamma_{f,(g,\gamma)}^{(4,1)}}{\Gamma_f^{(4,0)}} + 
 \frac{\overline{v}^2 }{\pi}\Gamma_f^{(4,0)} \nonumber \\ 
 	&
 	\times \bigg\{  \frac{m_H^2}{\sqrt{2}\overline{v} \overline{m}_f}
  \bigg(\delta_{f,q}\frac{C_F}{g_s} \alpha_s C_{fG}+ \frac{Q_f}{\overline{e}} \alpha
  \left( C_{fB} \hat{c}_w +2 T^3_f C_{fW} \hat{s}_w\right)\bigg) 
\nonumber \\ &
+\left(\delta_{f,q} C_F \alpha_s C_{HG}+ Q_f^2 \alpha\, c_{h\gamma\gamma}\right) 
  \left[19-\pi^2 + \ln^2\left(\frac{\overline{m}_f^2}{m_H^2}\right)+
  6\ln\left(\frac{\mu^2}{m_H^2}\right)\right] 
  \nonumber \\ &
  +c_{h\gamma Z} \,  v_f Q_f \alpha  \, F_{h\gamma Z}\left(\frac{M_Z^2}{m_H^2},\frac{\mu^2}{m_H^2},\frac{\overline{m}_f^2}{m_H^2}\right) 
  \bigg\} \, ,
 \end{align}
 where $v_f=(T^3_f-2 Q_f \hat{s}_w^2)/(2 \hat{s}_w \hat{c}_w)$ is the vector coupling of $f$ to the 
 $Z$-boson, $T^3_f$ is the weak isospin of fermion $f$ (i.e. $T^3_\tau=-\frac{1}{2}$ and $T^3_c=\frac{1}{2}$), $\delta_{f,q} =1$ if $f$ is a quark and $\delta_{f,q} =0$ if $f$ is a lepton, $C_F=(N_c^2-1)/(2N_c)$ with $N_c=3$, and the combinations of Wilson coefficients multiplying the $h\gamma\gamma$ and $h\gamma Z$ 
 vertices are
 \begin{align}
 \label{eq:WilsonVertex} 
 c_{h\gamma \gamma} &= C_{HB} \hat{c}_w^2 +C_{HW} \hat{s}_w^2-C_{HWB} \hat{c}_w \hat{s}_w \, , \nonumber \\
 c_{h\gamma Z} &= 2(C_{HB}-C_{HW})\hat{c}_w \hat{s}_w + C_{HWB}(\hat{c}_w^2-\hat{s}_w^2) \, . 
 \end{align}
In the small-mass limit the function $F_{h \gamma Z}$ takes the form
\begin{align}
& F_{h\gamma Z} \left(z, \hat{\mu}^2, 0 \right) =
-12+ 4z-\frac{4}{3} \pi^2 \bar{z}^2 +  \left(3+ 2z + 2\bar{z}^2 \ln (\bar{z})\right)\ln(z) 
+ 4 \bar{z}^2 {\rm Li}_2(z)  
	- 6\ln(\hat{\mu}^2) \,  .
\end{align}
where $\overline{z} = 1-z$. 

An interesting feature of Eq.~(\ref{eq:QEDQCDGam61}) is the double logarithm in the ratio $\overline{m}_f^2/m_H^2$ multiplying $C_{HG}$ and $c_{h\gamma\gamma}$.  In the SM, logarithms of this type first appear at NNLO, and are
related to diagrams where the Higgs couples to a top-quark loop which in the large-$m_t$ limit 
can be shrunk to an effective $hAA$ vertex,
where $A=\gamma,g$, multiplied by an $m_t$-dependent matching coefficient.  
These SM corrections, not only the logarithms but also the finite parts, are thus proportional
to the SMEFT corrections given above (see for instance Eq.~(8) of \cite{Chetyrkin:1996wr}).  As noted already in 
\cite{Larin:1995sq}, these double logarithms cancel against corresponding terms in the $h\to AA$ decay rate, 
such that total Higgs decay width remains finite in the limit of vanishing fermion masses.
In a less inclusive quantity such as $h\to ff$,  they introduce sizeable flavor-dependent contributions 
to the decay rate, even though  they multiply flavor-universal Wilson coefficients.  We return to this 
issue when studying ratios  of decay rates (in MFV) in Section~\ref{sec:Ratios}.

\subsection{Four-fermion operators}
\label{sec:4Fermion}

Contributions from four-fermion operators  (``class-8'' in Table~\ref{op59}) are obtained as in 
\cite{Gauld:2015lmb,  Cullen:2019nnr}.  
In this section we generalize those results to the generic decay $h\to f\bar{f}$, including contributions
from second-generation fermions which were neglected in previous analyses of $h\to b\bar{b}$ and 
$h\to \tau\bar{\tau}$.  

We first introduce some notation. In cases where top-quark loops contribute, the results involve the functions
\begin{align}
F_{8S}\left(\frac{m_t^2}{m_H^2},\frac{\mu^2}{m_H^2}\right)&=
\beta_t^2 \left(2\beta_t {\rm arccot}\left(\beta_t\right) - \ln\frac{\mu^2}{m_t^2}-2\right)\, , \nonumber  \\
F_{8V}\left(\frac{m_t^2}{m_H^2},\frac{\mu^2}{m_H^2}\right)&=
\beta_t^2 \left(-8\beta_t{\rm  arccot}\left(\beta_t\right)+4 \ln\frac{\mu^2}{m_t^2}+6\right)\, ,
\end{align}
where 
\begin{align}
\beta_t \equiv \sqrt{\frac{4m_t^2}{m_H^2}-1} \,.
\end{align}
Contributions from other fermions involve the real part of the above 
functions in the limit $m_t\to 0$, given by
\begin{align}
F_{8S}\left(0,\frac{\mu^2}{m_H^2}\right)&= 2+  \ln\frac{\mu^2}{m_H^2}  \, , \nonumber \\
F_{8V}\left(0,\frac{\mu^2}{m_H^2}\right)&=
-6- 4\ln\frac{\mu^2}{m_H^2}\, .
\end{align}
The functions with subscripts $8V$ arise from four-fermion operators of the form $(\bar{L}L)(\bar{R}R)$ in Table~\ref{op59},
whereas those with subscripts $8S$ arise from four-fermion operators of the form $(\bar{L}R)(\bar{R}L)$ or 
$(\bar{L}R)(\bar{L}R)$ given in the last row of that table.

In terms of the above functions, the results for $h\to \tau\bar{\tau}$ are
\begin{align}
\Gamma_{\tau 8} &= \frac{m_H^2 \Gamma^{(4,0)}_{\tau}}{16 \pi^2} \Bigg\{
\left[ C_{\substack{le \\ 3333}} +\frac{\overline{m}_\mu}{\overline{m}_\tau} C_{\substack{le \\ 2332}} \right] 
F_{8V}\left(0,\frac{\mu^2}{m_H^2}\right)-2N_c \frac{m_t}{\overline{m}_\tau} C^{(1)}_{\substack{lequ \\ 3333}}F_{8S}\left(\frac{m_t^2}{m_H^2},\frac{\mu^2}{m_H^2}\right) \nonumber \\
&+2 N_c \left[ \frac{\overline{m}_b}{\overline{m}_\tau} C_{\substack{ledq \\ 3333}}
- \frac{\overline{m}_c}{\overline{m}_\tau} C^{(1)}_{\substack{lequ \\ 3322}}  
+\frac{\overline{m}_s}{\overline{m}_\tau} C_{\substack{ledq \\ 3322}}\right]F_{8S}\left(0,\frac{\mu^2}{m_H^2}\right) 
\Bigg\} \,.
\end{align}
The expression for $h\to \mu\bar{\mu}$ is obtained after obvious replacements and reads
\begin{align}
\Gamma_{\mu 8} &= \frac{m_H^2 \Gamma^{(4,0)}_{\mu}}{16 \pi^2} \Bigg\{
\left[  C_{\substack{le \\ 2222}} + \frac{\overline{m}_\tau}{\overline{m}_\mu} C_{\substack{le \\ 2332}} \right] 
F_{8V}\left(0,\frac{\mu^2}{m_H^2}\right)-2N_c \frac{m_t}{\overline{m}_\mu} C^{(1)}_{\substack{lequ \\ 2233}}F_{8S}\left(\frac{m_t^2}{m_H^2},\frac{\mu^2}{m_H^2}\right) \nonumber \\
&+2 N_c \left[ \frac{\overline{m}_b}{\overline{m}_\mu} C_{\substack{ledq \\ 2233}}
- \frac{\overline{m}_c}{\overline{m}_\mu} C^{(1)}_{\substack{lequ \\ 2222}}  
+\frac{\overline{m}_s}{\overline{m}_\mu} C_{\substack{ledq \\ 2222}}\right]F_{8S}\left(0,\frac{\mu^2}{m_H^2}\right) 
\Bigg\} \,.
\end{align}
For $h\to c\bar{c}$, we find instead
\begin{align}
\Gamma_{c8}  &= \frac{m_H^2 \Gamma^{(4,0)}_{c}}{16 \pi^2} \Bigg\{ 
 \left[C^{(1)}_{\substack{qu \\ 2222}} + C_F C^{(8)}_{\substack{qu \\ 2222}} \right] F_{8V}\left(0,\frac{\mu^2}{m_H^2}\right)
\nonumber \\ &
+ \frac{m_t}{\overline{m}_c} \left[C^{(1)}_{\substack{qu \\ 2332}} +C_F C^{(8)}_{\substack{qu \\ 2332}} \right] 
F_{8V}\left(\frac{m_t^2}{m_H^2},\frac{\mu^2}{m_H^2}\right)+ 
\left( \frac{\overline{m}_b}{\overline{m}_c} \left[ C^{(1)}_{\substack{quqd \\ 3223}} +C_F C^{(8)}_{\substack{quqd \\ 3223}} + 2 N_c C^{(1)}_{\substack{quqd \\ 2233}} \right] \right. \nonumber \\ &
\left. + \frac{\overline{m}_s}{\overline{m}_c} \left[(1+ 2N_c) C^{(1)}_{\substack{quqd \\ 2222}} +C_F C^{(8)}_{\substack{quqd \\ 2222}}\right]
- 2 \frac{\overline{m}_\tau}{\overline{m}_c} C^{(1)}_{\substack{lequ \\ 3322}} - 2 \frac{\overline{m}_\mu}{\overline{m}_c} C^{(1)}_{\substack{lequ \\ 2222}}
\right)F_{8S}\left(0,\frac{\mu^2}{m_H^2}\right) 
\Bigg\} \, .
\end{align}
To simplify the results, we have used relations such as $C^{(k)}_{\substack{qu \\ 2332}}= C^{(k)\dagger}_{\substack{qu \\ 3223}}$ (with $k=1,8$) which follow from the Hermiticity of the SMEFT Lagrangian. 
Finally, the result for  $h\to b\bar{b}$ is 
\begin{align}
\Gamma_{b8}  &= \frac{m_H^2 \Gamma^{(4,0)}_{b}}{16 \pi^2} \Bigg\{
\left[C^{(1)}_{\substack{qd \\3333}} +C_F C^{(8)}_{\substack{qd \\ 3333}} + \frac{\overline{m}_s}{\overline{m}_b} \left(C^{(1)}_{\substack{qd \\ 2332}} +C_F C^{(8)}_{\substack{qd \\ 2332}} \right) \right] F_{8V}\left(0,\frac{\mu^2}{m_H^2}\right)
\nonumber \\ &
+\frac{m_t}{\overline{m}_b} \left[(1+2N_c) C^{(1)}_{\substack{quqd \\ 3333}} +C_F C^{(8)}_{\substack{quqd \\ 3333}} \right]
F_{8S}\left(\frac{m_t^2}{m_H^2},\frac{\mu^2}{m_H^2}\right) 
\nonumber \\ &
+ \left( \frac{\overline{m}_c}{\overline{m}_b} \left[ C^{(1)}_{\substack{quqd \\ 3223}} +C_F C^{(8)}_{\substack{quqd \\ 3223}} +2 N_c C^{(1)}_{\substack{quqd \\ 2233}} \right] + 2 \frac{\overline{m}_\tau}{\overline{m}_b}C_{\substack{ledq \\ 3333}} +2 \frac{\overline{m}_\mu}{\overline{m}_b} C_{\substack{ledq \\ 2233}} \right) F_{8S}\left(0,\frac{\mu^2}{m_H^2}\right)
\Bigg\} \, .
\end{align}
In all cases, contributions where the Higgs couples to a top-quark loop are enhanced by a large factor of $m_t/\overline{m}_f$.  If MFV is imposed, the Wilson coefficients multiplying these contributions pick up factors proportional to $\overline{m}_f$, 
thus removing  this enhancement.  Moreover,  contributions from top loops to decay into charm quarks vanish in MFV, under the approximation where the CKM matrix  is the unit matrix, as can be seen by using  MFV scalings for the 
Wilson coefficients in Section~\ref{sec:WilsonCoefficientsMFV}.

\section{Numerical results}
\label{sec:NumericalResults}
\begin{table}[t]
	\begin{center}
		\def\arraystretch{1.3}
		\begin{tabular}{|c|c||c|c|}
			\hline  $m_H$ &  $125$~GeV   &$\overline{e}(m_H)$ & $\sqrt{4 \pi/ 128}$ \\ 
			\hline $m_t$ & $173$~GeV& $\alpha_s \left(m_H\right)$ & 0.1 \\ 
			\hline $M_W$ & $80.4$~GeV &  $\overline{m}_b(m_H)$ & 3.0 GeV \\ 
			\hline $M_Z$ & $91.2$~GeV & $\overline{m}_\tau(m_H)$ & 1.7 GeV  \\
			\hline  $\overline{v}(m_H)$ & 240 GeV & $\overline{m}_c(m_H)$ & 0.7 GeV  \\
			\hline 
		\end{tabular} 
		\caption{\label{tab:NumericalTable} Input parameters used in numerics.  The
		 derived quantity $\overline{v}(m_H) = 2M_W \hat{s}_w/\overline{e}(m_H)$ is listed 
		for convenience.}
	\end{center}

\end{table}

In this section we give numerical results for the $h \to \tau \bar{\tau}$ and $h \to c \bar{c}$ decay rates.
We list contributions from the full set of Wilson coefficients at the scale $\mu=m_H$ in Section~\ref{sec:CentralValues},
and then examine scale uncertainties for the numerically dominant ones
in Section~\ref{sec:ScaleUncertainties}. The numerical inputs used throughout  the paper are given in Table~\ref{tab:NumericalTable}.

\subsection{Decay rates at  $\mu=m_H$}
\label{sec:CentralValues}
It is convenient to normalize all results to the LO SM decay rate at the scale $\mu=m_H$. 
To this end, in analogy with Eq.~(\ref{eq:PertExpansionSMEFT}) we define the ratios
\begin{align}
\label{eq:DeltaDef}
\Delta_f^{(i,j)}(\mu) \equiv \frac{\Gamma_f^{(i,j)}(\mu)}{\Gamma_f^{(4,0)}(m_H)} \, ,
\end{align}
as well as the LO and NLO ratios
\begin{align}
\label{eq:DefineDelta}
\Delta^{\rm LO}_{f}(\mu) &\equiv  \Delta_f^{(4,0)}(\mu) + \Delta_f^{(6,0)}(\mu)\, , 
 \nonumber \\
 \Delta^{\rm NLO}_f(\mu)& \equiv 
 \Delta^{\rm LO}_f(\mu) +   \Delta_f^{(4,1)}(\mu) + \Delta_f^{(6,1)}(\mu)  \, .
\end{align}
We also define dimensionless Wilson coefficients as
\begin{align}
  \tilde{C}_i(\mu)  \equiv  \Lambda_{\rm NP}^2 C_i(\mu)  \,,
\end{align} 
where $\Lambda_{\rm NP}$ is a UV scale characteristic of heavy new physics beyond the SM.
Contributions to the decay rate from dimension-6 operators are 
suppressed by a factor of  $\bar{v}^2/\Lambda_{\rm NP}^2 \approx 5\%$, where the numerical
value refers to $\Lambda_{\rm NP}=1$~TeV and $\tilde{C}_i\sim 1$.

The LO result for the decay $h\to f\bar{f}$ at the scale $\mu=m_H$ is
\begin{align}
\Delta^{\rm LO}_f&(m_H)  = 1 + \frac{\overline{v}^2}{\Lambda_{\rm NP}^2} \left[ 3.74 \Ct_{HWB}+2.00 \Ct_{H \Box}- 1.41 \frac{\overline{v}}{\overline{m}_{f}}\Ct_{f H} + 1.24 \Ct_{HD} \right] \, ,
\end{align}
where here and below all \msbar-renormalized quantities (in this case $\overline{v}$, $\tilde{C}_i$, $\overline{m}_f$) 
are evaluated at the scale $m_H$.   The contribution from $\Ct_{fH}$ is enhanced by a factor of 
$\overline{v}/\overline{m}_f$ compared to other contributions.  We have left this factor symbolic,
such that the numerical coefficient multiplying it is finite in the limit $m_f\to 0$.  In a theory which respects MFV,
the enhancement factor would indeed be compensated by an implicit scaling of the Wilson coefficient.
While we are not  necessarily advocating MFV, we choose to write all our results in such a way that all numerical
factors multiplying dimension-6 Wilson coefficients are free of large enhancement factors.  
This affects several more coefficients at NLO, and for the same reason we leave factors of 
$1/\overline{e}$ and $1/g_s$ appearing in operators such as $Q_{f(B,W)}$ and $Q_{fG}$, where
gauge fields couple to fermions through field-strength tensors rather than covariant derivatives, symbolic.

The NLO corrections depend on the flavor of the fermion $f$.  For $h\to\tau\bar{\tau}$, we find  
\begin{align}
\label{eq:CentralValueNLOtau}
\Delta^{\rm NLO}_\tau&(m_H)  = 0.98 + 
\frac{\overline{v}^2}{\Lambda_{\rm NP}^2}
\bigg\{ 3.63 \Ct_{HWB}+2.11 \Ct_{H \Box}- 1.50\frac{\overline{v}}{\overline{m}_{\tau}}\Ct_{\tau H} +1.20 \Ct_{HD} + 0.16 \Ct_{HB} \notag \\
	& + \Bigg(-9.0 \Ct^{(3)}_{\substack{Hq \\ 33}} -7.9 \Ct_{Ht} +6.8 \Ct_{HW} +5.2 \Ct^{(1)}_{\substack{Hq \\ 33}} -4.6 \Ct_{tH}  -3.8\frac{m_t}{\overline{m}_\tau} \Ct^{(1)}_{\substack{l e qu \\ 3333}} +2.4 \Ct_H   \nonumber \\
	&+2.0 \frac{\overline{m}_b}{\overline{m}_\tau}  \Ct_{\substack{ledq \\ 3333}} + 2.0 \frac{\overline{m}_s}{\overline{m}_\tau} \Ct_{\substack{ledq \\ 3322}} - 2.0 \frac{\overline{m}_c}{\overline{m}_\tau} \Ct^{(1)}_{\substack{lequ \\ 3322}} +1.5 \Ct^{(3)}_{\substack{Hl \\ 33}}-1.0 \Ct_{\substack{l e \\ 3333}} -1.0 \frac{\overline{m}_\mu}{\overline{m}_\tau} \Ct_{\substack{l e \\ 2332}} \Bigg)   \nonumber \\
	&\times 10^{-2}+ \Bigg(-9 \Bigg[ \frac{\Ct_{tB}}{\overline{e}}+ \Ct^{(3)}_{\substack{Hq \\ 11}} +\Ct^{(3)}_{\substack{Hq \\ 22}}  -\Ct^{(1)}_{\substack{Hl \\ 33}} +\Ct_{Hu}   +\Ct_{Hc} \Bigg]-7 \Ct_W+4 \bigg[ \Ct^{(1)}_{\substack{Hl \\ 11}}+ \Ct^{(1)}_{\substack{Hl \\ 22}} \nonumber \\
	&+\Ct_{He}+\Ct_{H\mu} - \Ct^{(1)}_{\substack{Hq \\ 11}}-\Ct^{(1)}_{\substack{Hq \\ 22}} + \Ct_{Hd} +\Ct_{Hs}+\Ct_{Hb} \bigg] -3 \left[\Ct^{(3)}_{\substack{Hl \\ 11}} +\Ct^{(3)}_{\substack{Hq \\ 22}} \right]+ 2 \frac{\Ct_{tW}}{\overline{e}}  \nonumber \\
	&+ \frac{\overline{v}}{\overline{m}_\tau} \frac{\Ct_{\tau W}}{\overline{e}} \Bigg) \times 10^{-3} + (4 \times 10^{-4})\left(  \Ct_{H \tau} - \frac{\overline{v}}{\overline{m}_\tau}\frac{\Ct_{\tau B}}{\overline{e}}  \right) \bigg\}\, .
\end{align}
While our main focus in this section is $h\to \tau\bar{\tau}$, we mention in passing that
results for $h\to \mu\bar{\mu}$ can be obtained from the above result by exchanging 
$\overline{m}_\tau \leftrightarrow \overline{m}_\mu$ along with appropriate changes on flavor indices of the 
Wilson coefficients.   The only implicit $m_\tau$ dependence in the small-mass limit is in the 
$\alpha \ln(\overline{m}_\tau^2/m_H^2)$ terms in Eq.~(\ref{eq:QEDQCDGam61}); for $h\to \mu\bar{\mu}$,
these change the coefficients of $\Ct_{HWB},\, \Ct_{HW},\, \Ct_{HB}$ from the values listed above to $3.49,\, 0.14, \, 0.41$, respectively.

The results for  $h\to c\bar{c}$ are given by
\begin{align}
\Delta^{\rm NLO}_c&(m_H)  = 1.16 + \frac{\overline{v}^2}{\Lambda_{\rm NP}^2}  \bigg\{ 4.95 \Ct_{HG}+ 4.31 \Ct_{HWB}+ 2.46  \Ct_{H \Box} - 1.75 \frac{\overline{v}}{\overline{m}_c} \Ct_{cH}  \nonumber \\
	&+  1.41 \Ct_{HD} + \Bigg(9.4 \Ct_{HB} -8.9 \Ct^{(3)}_{\substack{Hq \\ 33}} - 7.9 \Ct_{Ht} - 6.3 \frac{m_t}{\overline{m}_c} \Ct^{(8)}_{\substack{qu \\ 2332}} +  5.4 \Ct_{HW}   \nonumber \\ 
	& + 5.2 \Ct^{(1)}_{\substack{Hq \\33}} - 4.8 \frac{m_t}{\overline{m}_c} \Ct^{(1)}_{\substack{qu \\ 2332}} - 4.6 \Ct_{tH} +2.4 \Ct_H  +2.4 \frac{\overline{m}_s}{\overline{m}_c} \Ct^{(1)}_{\substack{quqd \\ 2222}} +2.0 \frac{\overline{m}_b}{\overline{m}_c} \Ct^{(1)}_{\substack{quqd \\ 2233}} - 1.3 \Ct^{(8)}_{\substack{qu \\ 2222}}   \nonumber \\
	& - 1.0 \Ct^{(1)}_{\substack{qu \\ 2222}}  + 1.0 \Ct^{(3)}_{\substack{Hq \\ 22}} - 1.0 \Ct^{(1)}_{\substack{Hq \\ 22}} \Bigg) \times 10^{-2} + \Bigg(-9 \left[ \frac{\Ct_{tB}}{ \overline{e}} +  \Ct^{(3)}_{\substack{Hq \\11}}+ \Ct_{Hu} \right] +8 \frac{\bar{v}}{\overline{m}_c} \frac{\Ct_{cG}}{g_s} \nonumber \\ 
	&-7 \frac{\overline{m}_\tau}{\overline{m}_c}\Ct^{(1)}_{\substack{lequ \\ 3322}}-7 \frac{\overline{m}_\mu}{\overline{m}_c} \Ct^{(1)}_{\substack{lequ \\ 2222}}-7 \Ct_{W}  - 5 \Ct_{Hc} + 4 \Bigg[\frac{\overline{m}_b}{\overline{m}_c} \Ct^{(8)}_{\substack{quqd \\ 3223}} + \frac{\overline{m}_s}{\overline{m}_c} \Ct^{(8)}_{\substack{quqd \\ 2222}} + \Ct^{(1)}_{\substack{Hl \\ 11}} + \Ct^{(1)}_{\substack{Hl \\ 22}}  \nonumber \\
	&+ \Ct^{(1)}_{\substack{Hl \\ 33}} - \Ct^{(1)}_{\substack{Hq \\ 11}} + \Ct_{He}   + \Ct_{H\mu} + \Ct_{H \tau} + \Ct_{Hd}+ \Ct_{Hs}+ \Ct_{Hb} \Bigg] - 3 \Big[-\frac{\overline{m}_b}{\overline{m}_c} \Ct^{(1)}_{\substack{quqd \\ 3223}} + \Ct^{(3)}_{\substack{Hl \\ 11}} \nonumber \\
	&+ \Ct^{(3)}_{\substack{Hl \\ 22}} + \Ct^{(3)}_{\substack{Hl \\ 33}}  \Big] + 2 \frac{\Ct_{tW}}{\overline{e}} + \frac{\overline{v}}{\overline{m}_c} \frac{\Ct_{cW}}{\overline{e} } \Bigg) \times 10^{-3}+ (2\times 10^{-4}) \frac{\overline{v}}{ \overline{m}_c } \frac{\Ct_{cB} }{\overline{e}} \bigg\}  \, .
\end{align}

Compared to LO, a total of 36 new Wilson coefficients appear at NLO in $h\to \tau\bar{\tau}$ and 43 in $h \to c\bar{c}$. 
While the size of these contributions depends on the values of the in general unknown Wilson coefficients, it is possible
to make some general statements.  

First, in all cases the potentially largest NLO contributions are from those coefficients which carry a non-trivial scaling with 
$\overline{m}_f$ in MFV.  Higgs decays thus offer an interesting testing ground for MFV, 
which becomes more involved beyond LO.  Second, since leptonic decays $h\to \ell\bar\ell$ have no QCD corrections to this 
order, the NLO corrections are generally mild, the most sizeable (apart from MFV-violating enhancements) being
that from $\Ct_{HB}$, which is enhanced by the double logarithm in $\overline{m}_\ell/m_H$ in 
Eq.~(\ref{eq:QEDQCDGam61}).  For $h\to c\bar{c}$ this double logarithm also multiplies the QCD correction from $\Ct_{HG}$,
and the correction is so large that actually dominates over coefficients appearing at LO.  Finally, the corrections to the operators
appearing at LO are much larger in $h\to c\bar{c}$ than in $h\to\ell\bar{\ell}$ due to QCD corrections, although in neither
case are these well approximated by applying an SM-like $K$-factor to the LO results.  To illustrate this latter point we show
in Table~\ref{tab:CorSplitM} the NLO corrections to coefficients appearing at LO in $h\to\tau\bar{\tau}$ and $h\to c\bar{c}$ decays,  split into the QED-QCD and weak corrections, i.e.  $\Gamma^{(i,1)}_{f, (g,\gamma)}$ and  $\Gamma_{f,\rm weak}^{(i,1)}$ in Eq.~(\ref{eq:PertExpansion2}), respectively.

\begin{table}[t]
\begin{center}
\begin{tabular}{|l|rrrrr|}
\hline
$h\to \tau\bar\tau$ & SM  & $\Ct_{HWB}$
& $\Ct_{H\Box}$  & $\Ct_{f H}$  & $\Ct_{HD}$ \\
\hline 
NLO QED  & 1.1\% & -1.3 \% & 1.1\%  & 1.1\% &  1.1\%   \\
NLO weak & -2.8\% & -1.5\% &  4.5\%  & 5.3\% &  -5.0\%   \\
\hline 
NLO correction & -1.7\% & -2.9\% &  5.5\%  & 6.3\% &  -3.9\%   \\
\hline
$h \to c\bar{c}$ &&&&&\\ \hline 
NLO QCD-QED  & 18.5\% & 17.0 \% & 18.5\%  & 18.5\% &  18.5\%   \\
NLO weak & -2.8\% & -1.6 \% &  4.4\%  & 5.3\% &  -5.2\%   \\\hline 
NLO correction & 15.7\% & 15.3 \% &  22.9 \%  & 23.8\% &  13.3\%   \\ 
\hline
\end{tabular}
\caption{\label{tab:CorSplitM} 
Size of NLO corrections to different terms in LO decay rate, split into
QED(-QCD) and weak corrections for  $h \to \tau \bar{\tau}$ (top) and $h \to c \bar{c}$ (bottom). 
See text for further explanation.  }
\end{center}
\end{table}

\subsection{Scale uncertainties}
\label{sec:ScaleUncertainties}
The truncation of the perturbative series to finite order introduces dependence of the decay rates on 
unphysical renormalization scales.  The change in the decay rate under ranges of scale choices is often
used to estimate uncertainties associated with uncalculated, higher-order corrections.  Ideally, the uncertainty bands
shrink upon adding higher-order perturbative corrections, and also show good overlap between orders.   

In this section, we study scale uncertainties in the LO and NLO SMEFT calculations of $h\to \tau\bar{\tau}$
and $h\to c\bar{c}$ decay rates.  This requires knowledge of how the \msbar-renormalized input 
parameters depend on renormalization scale.  In the SM this is simple, as the light-fermion masses and  
QED/QCD couplings have been extracted numerically at  certain reference scales, and can be evolved 
to any other scale using RG equations.  

In SMEFT, on the other hand, the dimension-6 Wilson coefficients have no accepted numerical values.  In order to study scale uncertainties in a meaningful way, we must therefore express decay rates at different scales in terms of Wilson coefficients evaluated at a fixed reference scale.  We choose this reference scale to be  $m_H$. 
The Wilson coefficients at an arbitrary scale $\mu_C$ can then 
be determined from their values at $m_H$ through use of RG equations.  Since in our analysis the $\mu_C$ will be at most 
a factor of two away from the reference value $\mu_C = m_H$, 
we can use the fixed-order solutions to the RG equations.  In fact, the same
is true of the \msbar-renormalized masses and couplings,  so to study scale variations we will 
need only the following equations: 
\begin{align}
\label{eq:Cevolve1}
C_i(\mu_C) & = C_i(m_H) +\ln\left(\frac{\mu_C}{m_H}\right)
 \dot{C}_i(m_H) \,  ,
\nonumber \\
\overline{m}_f(\mu_R)&=\overline{m}_f(m_H)
\left[1+ \gamma_f(m_H) \ln\left(\frac{\mu_R}{m_H}\right)\right] ,
\nonumber \\
\overline{\alpha}(\mu_R) & = \overline{\alpha}(m_H) \left[ 1+2 \gamma_e(m_H) \ln\left(\frac{\mu_R}{m_H}\right) \right] \, , \nonumber \\
\alpha_s(\mu_R) & = \alpha_s(m_H)\left[ 1-2 \gamma_g(m_H) \ln\left(\frac{\mu_R}{m_H}\right) \right] \, .
\end{align}
Expressions for $\dot{C}_i \equiv dC_i/d\ln\mu$ at one loop can be found in 
\cite{Jenkins:2013zja, Jenkins:2013wua, Alonso:2013hga}, and we have made use of their
electronic implementation from  \cite{Celis:2017hod}. The corresponding SM expressions are 
\begin{align}
\gamma_f(\mu) &=-\frac{3}{2\pi}\left[ \delta_{f,q} \alpha_s(\mu) C_F + \overline{\alpha}(\mu) Q_f^2\right]  \, , 
\nonumber \\
\gamma_e(\mu) & = 
\frac{\overline{\alpha}(\mu)}{3\pi}\left[ 3 Q_\ell^2 +N_c \left(2 Q_u^2+3 Q_b^2\right) \right]\,, \nonumber  \\ 
\gamma_g(\mu) &= \frac{\alpha_s(\mu)}{4\pi} \left( \frac{11}{3} C_A-\frac{2}{3}n_l\right)   \, ,
\end{align}
where $n_l=5$ is the number of light fermions and $C_A=3$.  The notation makes clear that we are free
to choose the two renormalization scales $\mu_C$ and $\mu_R$ independently. We will indeed do so, 
such that the expansion coefficients in Eq.~(\ref{eq:DeltaDef}) become functions of two scales,  $\Delta_f^{(i,j)}(\mu)\to \Delta_f^{(i,j)}(\mu_C,\mu_R)$.  The explicit logarithmic dependence on the two scales  can be deduced 
through an RG analysis and is given in  Eq.~(5.7) of \cite{Cullen:2019nnr}. 

With these expressions at hand, we can fix a procedure for quantifying scale uncertainties.  
We follow the method advocated in \cite{Cullen:2019nnr}.  First, we set $\mu_C=\mu_R=m_H$ by 
default.  We then vary $\mu_C$ to the values $m_H/2$ and $2m_H$, keeping $\mu_R=m_H$, and use the results
to calculate upper and lower uncertainties from $\mu_C$ variations.  We obtain uncertainties from $\mu_R$
variations analogously, and add the $\mu_C$ and $\mu_R$ uncertainties in quadrature to get a total uncertainty.  

For $h\to \tau\bar{\tau}$, the LO result obtained in this way is
\begin{align}
\label{eq:LOtauScaleVar}
\Delta^{\rm LO}_\tau&(m_H, m_H)  = (1^{+0.002}_{-0.003}) + \frac{\overline{v}^2}{\Lambda_{\rm NP}^2} \Big\{ (3.74 \pm 0.14) \Ct_{HWB}+ (2.00 \pm 0.12) \Ct_{H \Box}  \nonumber \\
	& - (1.41 \pm 0.06) \frac{\overline{v}}{\overline{m}_{\tau}}\Ct_{\tau H} + (1.24 \pm 0.09) \Ct_{HD} \pm 0.19 \Ct^{(1)}_{\substack{Hq \\ 33}} \pm 0.18\Ct_{Ht} \pm 0.09 \frac{m_t}{\overline{m}_\tau}\Ct^{(1)}_{\substack{lequ \\ 3333}} \nonumber \\
	& \pm 0.05\Ct_{tH}  \pm  0.05 \Ct^{(3)}_{\substack{Hq \\ 33}} 
	+... \Big\} \, ,
\end{align}
while at NLO we find
\begin{align}
\Delta^{\rm NLO}_\tau &(m_H, m_H)  = (0.98^{+0.0001}_{-0.0002}) + \frac{\overline{v}^2}{\Lambda_{\rm NP}^2} \Big\{ (3.62^{+0.00}_{-0.01}) \Ct_{HWB}+(2.11^{+0.00}_{-0.02}) \Ct_{H \Box} \nonumber \\
	&+ (-1.50^{+0.01}_{-0.00}) \frac{\overline{v}}{\overline{m}_\tau} \Ct_{\tau H} + (1.20^{+0.00}_{-0.01}) \Ct_{HD}+(0.16^{+0.00}_{-0.00}) \Ct_{HB} + (-0.09^{+0.02}_{-0.00}) \Ct^{(3)}_{\substack{Hq \\ 33}}  \nonumber \\
	&+ (-0.08^{+0.03}_{-0.00})  \Ct_{Ht} +  (0.07^{+0.00}_{-0.00}) \Ct_{HW} + (0.05^{+0.00}_{-0.03} ) \Ct^{(1)}_{\substack{Hq \\ 33}}+ (-0.05^{+0.01}_{-0.00}) \Ct_{tH} 
 + ... \Big\} \, .
\end{align}
In the above equations and in the results for $h\to c\bar{c}$ that follow, the ellipses denote terms whose magnitude is smaller
than 5\%.\footnote{More precisely, for a term of the form $x^{+y}_{-z}$, we neglect terms where both
$|x+y|$ and $|x-z|$ are smaller than 5\%.}  

The uncertainties are significantly reduced when including the NLO corrections, and in general the LO
uncertainty ranges encompass the true value of the NLO correction, showing a good convergence
of the perturbative series.  There are two notable exceptions.  The first is the SM itself, where the LO and NLO uncertainty
bands show little overlap.  This is mainly due to the fact that there is a correlation between the running of $\overline{m}_\tau$
and $\overline{v}$, such that the ratio $\overline{m}_f(\mu_R)/\overline{v}(\mu_R)$ is more stable under scale variations
that the numerator and denominator alone.  A more conservative option would be to introduce and vary independently 
separate renormalization scales for the fermion masses and the electromagnetic coupling constant, 
but we do not pursue that option here. The second exception is $C_{HB}$ and $C_{HW}$, which get NLO corrections proportional to $\alpha \ln^2(\overline{m}_\tau^2/m_H^2)$ that are inaccessible to an RG analysis.   

For $h\to c\bar{c}$, the LO result is 
\begin{align}
\Delta^{\rm LO}_c &(m_H,m_H)  = (1 \pm 0.08) + \frac{\overline{v}^2}{\Lambda_{\rm NP}^2} \Big\{ 
(3.74^{+0.37}_{-0.36}) \Ct_{HWB} + (2.00^{+0.22}_{-0.21}) \Ct_{H \Box} \nonumber \\
&+ (-1.41 \pm 0.07)  \frac{\overline{v}}{\overline{m}_{c}}\Ct_{c H} + (1.24^{+0.15}_{-0.14}) \Ct_{HD} 
 \pm 0.35 \Ct_{HG} 
 \pm 0.19 \Ct_{\substack{Hq \\ 33}}^{(1)} 
   \nonumber \\ 
&\pm 0.18 \Ct_{Ht} \pm 0.05 \Ct_{tH} \pm 0.05 \Ct_{\substack{Hq \\ 33}}^{(3)} 
	+ ... \Big\} \, ,
\end{align}
and at NLO we have
\begin{align}
\Delta^{\rm NLO}_c &(m_H,m_H)  = (1.16^{+0.02}_{-0.04}) + \frac{\overline{v}^2}{\Lambda_{\rm NP}^2} \Big\{ (4.95^{+0.89}_{-0.84}) \Ct_{HG}+(4.31^{+0.06}_{-0.15}) \Ct_{HWB} \nonumber \\ 
	& + (2.46^{+0.05}_{-0.10}) \Ct_{H \Box} + (-1.75^{+0.04}_{-0.03})  \frac{\overline{v}}{\overline{m}_{c}}\Ct_{c H} + (1.41^{+0.02}_{-0.05}) \Ct_{HD} +(0.00^{+0.12}_{-0.08}) \frac{\Ct_{tG}}{g_s}  \nonumber \\
	& +(0.09^{+0.01}_{-0.01}) \Ct_{HB} +(-0.09^{+0.03}_{-0.01}) \Ct_{\substack{Hq \\ 33}}^{(3)} + (-0.08^{+0.06}_{-0.01}) \Ct_{Ht} +(-0.06^{+0.08}_{-0.08})\frac{m_t}{\overline{m}_c} \Ct^{(8)}_{\substack{qu \\ 2332}}  \nonumber \\
	&+ (0.05^{+0.01}_{-0.01}) \Ct_{HW} +(0.05^{+0.01}_{-0.06}) \Ct^{(1)}_{\substack{Hq \\ 33}} + (-0.05^{+0.06}_{-0.06})\frac{m_t}{\overline{m}_c} \Ct^{(1)}_{\substack{qu \\ 2332}} + (-0.05^{+0.02}_{-0.02}) \Ct_{tH}
+ ... \Big\} \, .
\end{align}
The qualitative behavior is similar to the case of $h\to\tau\bar{\tau}$. There are however some differences for decay into quarks
rather than leptons.  There is a worse overlap between the LO and NLO results for the SM and dimension-6 coefficients appearing at tree level.  This is mainly due to the fact that the NLO QCD corrections in the SM are slightly larger than
what one gets by varying the scale in $\overline{m}_c$.  More significantly, the correction proportional to
$\alpha_s \ln^2(\overline{m}_c^2/m_H^2)$ multiplying the $C_{HG}$ contribution is much larger than the
corresponding terms in $h\to \tau\bar{\tau}$, due to the appearance of the QCD coupling, and is well outside the 
LO uncertainty estimate.  In fact,  the running of  $C_{HG}$ introduces a new coefficient, $C_{tG}$,
into the NLO uncertainty estimate, which is an $\alpha_s^2 \ln^2(\overline{m}_c^2/m_H^2)$ correction to the 
LO result and is numerically similar to the  NLO corrections for other Wilson coefficients.   

Since $C_{HG}$ itself gets NNLO corrections of order $\alpha_s^2 \ln^4(\overline{m}_c^2/m_H^2)$, which are  
substantial, it is clear that more reliable predictions for $h\to c\bar{c}$ would involve a resummation of the 
logarithmic terms. While techniques for such a resummation exist for the virtual $h\to AA$ amplitudes (where $A=\gamma,g$) 
\cite{Akhoury:2001mz, Liu:2017vkm,  Liu:2018czl, Liu:2019oav, Wang:2019mym}, 
it is not clear how to translate them to the inclusive $h\to f\bar{f}$ decay rate, which receives double-logarithmic
corrections from both real and virtual contributions.  

\section{Ratios of decay rates}
\label{sec:Ratios}

LHC measurements are to a large extent limited to $h\to b\bar{b}$ and $h\to \tau \bar\tau$.  At lepton 
colliders, on the other hand, measurements of Higgs decays into muons and charm quarks should 
also be possible, which motivates studying ratios of decay rates into different fermions.  This is a benefit 
in the SM, as ratios of decay rates are more stable under perturbative corrections than the decay
rates themselves.  It is even more advantageous in SMEFT,  as many flavor-universal dimension-6 contributions 
drop out of ratios.  This also eliminates, to a large extent, dependence on the choice of input parameters used in the 
renormalization procedure, for instance on the freedom to trade $\overline\alpha$ for the Fermi constant 
$G_F$ as is often done in SM calculations.  Finally, as we shall see in a moment, ratios of decay rates
offer interesting tests of MFV, as well as probes of anomalous $hgg$ and $h\gamma\gamma$ couplings induced
by SMEFT operators.

From the theoretical side, the cleanest possible ratios are those involving fermions with the same charges under the 
SM gauge group.  Lepton colliders should be able to measure the tau to muon ratio, so we focus
on this as a concrete possibility.  Let us first define
\begin{align}
\label{eq:Rtaumu}
R_{\tau/\mu}=\frac{\overline{m}_\mu^2}{\overline{m}_\tau^2}\frac{\Gamma_\tau}{\Gamma_\mu} \,.
\end{align}
In the SM,  up to NLO in perturbation theory and ignoring terms suppressed by $m_\ell^2/m_H^2$, $R_{\tau/\mu}=1$.  
In SMEFT, keeping only dimension-6 terms in the ratio up to NLO in perturbation theory and using 
$\overline{m}_\mu(m_H)=100$~MeV, we find the following  result at the scale $\mu=m_H$:
\begin{align}
\label{eq:RtaumuNum}
R_{\tau/\mu} -1 &= \Delta^{\rm NLO}_\tau(m_H)- \Delta^{\rm NLO}_\mu(m_H) - \Delta^{(4,1)}_\tau\left(\Delta^{(6,0)}_\tau - \Delta^{(6,0)}_\mu \right) \nonumber \\
&= \frac{\overline{v}^2}{\Lambda^2_{\rm NP}} \Bigg\{-1.53 \Bigg(\frac{\overline{v}}{\overline{m}_\tau} \Ct_{\tau H} - \frac{\overline{v}}{\overline{m}_\mu} \Ct_{\mu H} \Bigg)  - 0.25 \Ct_{HB} + 0.13 \Ct_{HWB} +\Bigg[-7.2 \Ct_{HW}  \nonumber \\
	&- 3.8 \Bigg( \frac{m_t}{\overline{m}_\tau} \Ct^{(1)}_{\substack{lequ \\ 3333}} - \frac{m_t}{\overline{m}_\mu} \Ct^{(1)}_{\substack{lequ \\ 2233}} \Bigg) 
	+    2.0 \Bigg(\frac{\overline{m}_b}{\overline{m}_\tau} \Ct_{\substack{ledq \\ 3333}}   
	- \frac{\overline{m}_b}{\overline{m}_\mu} \Ct_{\substack{ledq \\ 2233}} 
	+ \frac{\overline{m}_s}{\overline{m}_\tau} \Ct_{\substack{ledq \\ 3322}} 
	- \frac{\overline{m}_s}{\overline{m}_\mu} \Ct_{\substack{ledq \\ 2222}}  
		\nonumber \\
	& - \frac{\overline{m}_c}{\overline{m}_\tau} \Ct^{(1)}_{\substack{lequ \\ 3322}} + \frac{\overline{m}_c}{\overline{m}_\mu} \Ct^{(1)}_{\substack{lequ \\ 2222}}  \Bigg) + 1.8 \Bigg( \Ct^{(3)}_{\substack{Hl \\ 33}} - \Ct^{(3)}_{\substack{Hl \\ 22}} \Bigg) - 1.0 \Bigg( \left( \frac{\overline{m}_\mu}{\overline{m}_\tau} - \frac{\overline{m}_\tau}{\overline{m}_\mu} \right) \Ct_{\substack{le \\ 2332}}  + \Ct_{\substack{le \\ 3333}}  \nonumber \\
	& - \Ct_{\substack{le \\ 2222}} \Bigg) \Bigg] \times 10^{-2} + \Bigg[ 4 \Bigg( \Ct^{(1)}_{\substack{Hl \\ 33}} - \Ct^{(1)}_{\substack{Hl \\ 22}} - \Ct_{H \tau} + \Ct_{H \mu} \Bigg)  + \Bigg( \frac{\overline{v}}{\overline{m}_\tau} \frac{\Ct_{\tau W}}{\overline{e} }   \nonumber \\
	&- \frac{\overline{v}}{\overline{m}_\mu} \frac{\Ct_{\mu W}}{\overline{e} } \Bigg) \Bigg] \times 10^{-3} - (4 \times 10^{-4}) \Bigg( \frac{\overline{v}}{ \overline{m}_\tau}\frac{ \Ct_{\tau B} }{\overline{e}} - \frac{\overline{v}}{\overline{m}_\mu}\frac{ \Ct_{\mu B} }{\overline{e} }\Bigg) \Bigg\} \, ,
\end{align}
where have used that $\Delta_{\mu}^{(4,1)}=\Delta_{\tau}^{(4,1)}$ in the first equality.  Whereas $\Gamma_\tau+\Gamma_\mu$  depends on 48 Wilson coefficients, $R_{\tau/\mu}$ depends on 26 Wilson coefficients, as almost all operators that do not contain fermion fields drop out of the ratio.    

The results simplify if some form of universality is assumed for the generation-dependent Wilson 
coefficients.  As a concrete realization, we study the scenario where the SMEFT Wilson coefficients
are constrained by MFV.   The expressions and notation needed in the analysis are given in Section~\ref{sec:WilsonCoefficientsMFV}.   Dropping contributions to the ratio that are  suppressed by powers of 
$\overline{m}_f/m_H$ after taking the MFV limit,  we find the simple result 
\begin{align}
\bigg[R_{\tau/\mu} -1 \bigg]_{\rm MFV}  & = \frac{\alpha}{\pi} \mathcal{C}_{h\gamma\gamma}
 \left(\ln^2\frac{\overline{m}_\tau^2}{m_H^2} -\ln^2\frac{\overline{m}_\mu^2}{m_H^2} \right)
\nonumber \\ 
& =  \frac{\overline{v}^2}{\Lambda^2_{\rm NP}} \left( - 0.25 \mCt^0_{HB} + 0.13 \mCt^0_{HWB} - 0.072 \mCt^0_{HW}  \right) \,,
\end{align}
where $\mathcal{C}_{h \gamma \gamma}$ has been obtained from $c_{h \gamma \gamma}$
in Eq.~(\ref{eq:WilsonVertex}) by 
replacing the Wilson coefficients by their LO expansion in the small-mass limit as explained at the end of Section~\ref{sec:WilsonCoefficientsMFV}.
We have listed the analytic expression in order to emphasize that the deviation of the ratio from 
unity is due to the double logarithmic  corrections generated by the 
effective $h\gamma\gamma$ vertex and given explicitly in  Eq.~(\ref{eq:QEDQCDGam61}).
Uncertainties from higher orders estimated from scale variations using
the method described in the previous section are found to be less than 1\% for all dimension-6 coefficients involved.

As a second example, involving quarks instead of leptons, we consider the ratio $R_{c/b}$, defined in analogy to Eq.~(\ref{eq:Rtaumu}).  The result is not as simple as for $R_{\tau/\mu}$, since the $b$ and $c$ quarks have different
couplings to the electroweak gauge bosons.   QCD corrections cancel from the SM result up to NLO, so deviations
from unity are due to both to SM electroweak corrections and dimension-6 effects.   In particular, we find
\begin{align}
\label{eq:RcbNum}
R_{c/b} - 1 &= \Delta^{\rm NLO}_c - \Delta^{\rm NLO}_b - \Delta_c^{(4,1)} \Delta_b^{(6,0)} - \Delta^{(4,1)}_b \Delta^{(6,0)}_c + 2 \Delta_b^{(4,1)} \Delta_b^{(6,0)} \nonumber \\
& = 0.03+ \frac{\overline{v}^2}{\Lambda^2_{\rm NP}} \Bigg\{ 2.20 \Ct_{HG} - 1.57 \frac{\overline{v}}{\overline{m}_c} \Ct_{cH} + 1.59 \frac{\overline{v}}{\overline{m}_b} \Ct_{bH} +\Bigg[8.5 \Ct_{HB} - 6.3 \frac{m_t}{\overline{m}_c} \Ct^{(8)}_{\substack{qu \\ 2332}} \nonumber \\
	& -4.8 \frac{m_t}{\overline{m}_c} \Ct^{(1)}_{\substack{qu \\ 2332}}- 4.4 \frac{m_t}{\overline{m}_b} \Ct^{(1)}_{\substack{quqd \\ 3333}} +4.4 \Ct_{HWB}  + 4.2 \Ct_{HD}  + 2.7 \Ct_{HW} +2.6 \Ct^{(3)}_{\substack{Hq \\ 33}} \nonumber \\
	& +2.4 \frac{\overline{m}_s}{\overline{m}_c} \Ct^{(1)}_{\substack{quqd \\ 2222}} + 2.0 \left( \frac{\overline{m}_b}{\overline{m}_c} -\frac{\overline{m}_c}{\overline{m}_b} \right) \Ct^{(1)}_{\substack{quqd \\ 2233}} + 1.9 \Ct^{(3)}_{\substack{Hq \\ 22}} + 1.9 \frac{\overline{v}}{\overline{m}_b}\frac{ \Ct_{bW}}{\overline{e} } - 1.5 \Ct_{tH} +1.5 \frac{\Ct_{tW}}{\overline{e}}  \nonumber \\
	&- 1.3 \Bigg( \Ct^{(8)}_{\substack{qu \\ 2222}} - \Ct^{(8)}_{\substack{qd \\ 3333}} - \frac{\overline{m}_s}{\overline{m}_b} \Ct^{(8)}_{\substack{qd \\ 2332}} \Bigg) - 1.0 \Bigg( \Ct^{(1)}_{\substack{qu \\ 2222}} - \Ct^{(1)}_{\substack{qd \\ 3333}} - \frac{\overline{m}_s}{\overline{m}_b} \Ct^{(1)}_{\substack{qd \\ 2332}} \Bigg) \Bigg] \times 10^{-2}  \nonumber \\
	&+ \Bigg[ 8 \bigg( \frac{\overline{v}}{ \overline{m}_c}\frac{ \Ct_{cG}}{g_s} - \frac{\overline{v}}{ \overline{m}_b}\frac{ \Ct_{bG} }{g_s}\bigg) - 8 \frac{m_t}{\overline{m}_b} \Ct^{(8)}_{\substack{quqd \\ 3333}} - 7 \Bigg( \frac{\overline{m}_\tau}{\overline{m}_c} \Ct^{(1)}_{\substack{lequ \\ 3322}}+\frac{\overline{m}_\tau}{\overline{m}_b} C_{\substack{ledq \\ 3333}}  + \frac{\overline{m}_\mu}{\overline{m}_c} \Ct^{(1)}_{\substack{lequ \\ 2222}}   \nonumber \\
	&+ \frac{\overline{m}_\mu}{\overline{m}_b} C_{\substack{ledq \\ 2233}}  \Bigg) - 6 \Ct^{(1)}_{\substack{Hq \\ 33}}  -5 \Ct^{(1)}_{\substack{Hq \\ 22}} +4 \left( \frac{\overline{m}_b}{\overline{m}_c} - \frac{\overline{m}_c}{\overline{m}_b} \right) \Ct^{(8)}_{\substack{quqd \\ 3223}} + 4 \frac{\overline{m}_s}{\overline{m}_c} \Ct^{(8)}_{\substack{quqd \\ 2222}} + 4 \frac{\overline{v}}{\overline{m}_b} \Ct_{Htb}  \nonumber \\
	&+ 3 \left( \frac{\overline{m}_b}{\overline{m}_c} - \frac{\overline{m}_c}{\overline{m}_b} \right) \Ct^{(1)}_{\substack{quqd \\ 3223}} + 3 \Ct_{Hc} + 2\Ct_{Hb} + \frac{\overline{v}}{\overline{m}_c}\frac{ \Ct_{cW}}{\overline{e}} \Bigg] \times 10^{-3} + (2 \times 10^{-4}) \frac{\overline{v}}{ \overline{m}_c} \frac{\Ct_{cB}}{\overline{e}} \nonumber \\
	&+ (4 \times 10^{-5}) \frac{\overline{v}}{\overline{m}_b} \frac{\Ct_{bB} }{\overline{e}}\Bigg\} \, .
\end{align}
The deviation of the ratio from unity is only 3\% in the SM, and while
$\Gamma_b+\Gamma_c$  depends on 60 Wilson coefficients, $R_{c/b}$ depends on 41 Wilson coefficients. 
Again imposing MFV we find
\begin{align}
\bigg[R_{c/b} -1 \bigg]_{\rm MFV} &= 0.03 + \frac{\overline{v}^2}{\Lambda_{\rm NP}^2} \Bigg\{2.24 \mCt^1_{bH} - 2.22 \mCt^1_{cH} + 2.20 \mCt^0_{HG} + \Bigg[ 8.5 \mCt^0_{HB}-4.5 \mCt^{(1),2}_{\substack{quqd \\ 3333}}    \nonumber \\
	&+4.4 \mCt^0_{HWB} +4.2 \mCt^0_{HD} + 2.7 \frac{\mCt^1_{bW}}{\overline{e}} + 2.7 \mCt^0_{HW} +2.6 \mCt^{(3),0}_{\substack{Hq \\33}} +1.9 \mCt^{(3),0}_{\substack{Hq \\ 22}} - 1.6 \mCt^1_{tH}  \nonumber \\
	&+1.5 \frac{\mCt^1_{tW}}{\overline{e}} - 1.3 \left(\mCt^{(8),0}_{\substack{qu \\ 2222}}- \mCt^{(8),0}_{\substack{qd \\ 33}} \right) + 1.1 \left( \frac{\mCt^1_{cG}}{g_s} - \frac{\mCt^1_{bG}}{g_s} \right) \nonumber \\
	& -1.0 \left(\mCt^{(1),0}_{\substack{qu \\ 2222}}-\mCt^{(1),0}_{\substack{qd \\ 33}} \right)  \Bigg] \times 10^{-2} + \Bigg[ -9 \mCt^{(8),2}_{\substack{quqd \\ 3333}} -6 \mCt^{(1),0}_{\substack{Hq \\ 33}} -5 \mCt^{(1),0}_{\substack{Hq \\ 22}} +5 \mCt^2_{Htb}  \nonumber \\
	&+ 3 \mCt^0_{Hc}+ 2 \mCt^0_{Hb}+ 2 \frac{\mCt^1_{cW}}{\overline{e}} \Bigg] \times 10^{-3} + (3 \times 10^{-4}) \frac{\mCt^1_{cB}}{\overline{e}} + (6 \times 10^{-5}) \frac{\mCt^1_{b B}}{\overline{e}} \Bigg\} \, .
\end{align}
where we have used that $\mathcal{C}^{(1,3),0}_{\substack{Hq \\ 11}}=\mathcal{C}^{(1,3),0}_{\substack{Hq \\ 22}}$.
The result still depends on 28 dimension-6 Wilson coefficients, a number which can only be reduced by assuming
universality for Wilson coefficients of operators containing up and down type quarks of different generations 
not present in MFV.  Still, regardless of any flavor assumptions by far the largest numerical contributions 
multiply $\mCt_{HG}^0$,  $\mCt^1_{bH}$, and $ \mCt^1_{cH}$. Contributions from the additional coefficients $\mCt^0_{HD}$, $\mCt^0_{HWB}$, $\mCt^0_{H\Box}$, which appear at LO in the decay rates, 
either begin at NLO in the ratio, or else drop out entirely.  The contribution from $\mCt^0_{HG}$ is due to terms of the form
$\alpha_s \ln^2(\overline{m}_f^2/m_H^2)$, so as remarked above it would be desirable to resum such terms to obtain
a more reliable prediction.

\section{Conclusions}
\label{sec:Conclusions}

We have computed the NLO corrections from dimension-6 SMEFT operators to the decays $h\to f\bar{f}$, where
$f\in\{\mu,\tau, c\}$, thus extending the results for $h\to b\bar{b}$ from \cite{Cullen:2019nnr} to cover the
full spectrum of phenomenologically viable Higgs decays into fermions.  

Many lessons learned from $h\to b\bar{b}$ carry over to the generic decay modes $h\to f\bar{f}$. 
For instance, it is advantageous to use a  hybrid renormalisation scheme which avoids spuriously large tadpole corrections while resumming a series of single-logarithmic corrections in the ratio $\overline{m}_f/m_H$ through the use of running masses $\overline{m}_f$, defined in five-flavor version of QED$\times$QCD.  At the same time, it is a poor approximation to estimate NLO corrections to the dimension-6 operators appearing in the LO result by assuming they are proportional to the NLO
correction in the SM. 

On the other hand, considering several modes at once also opens up the possibility to study ratios of decay modes,
which turns out to be even more advantageous in SMEFT than in the SM, since many flavor-universal dimension-6 
Wilson coefficients drop out of the ratios.  Amusingly, even in scenarios where the Wilson coefficients are constrained 
by MFV, the flavor-universal SMEFT coefficients which alter the $hgg$ and $h\gamma\gamma$ couplings introduce
at NLO flavor-dependence in ratios of decay rates normalised to their respective LO SM results.  This dependence is from
double logarithmic corrections of the form $(\alpha,\alpha_s)\ln^2(\overline{m}_f^2/m_H^2)$, which cancel in total 
decay width but are present in the exclusive decay into a particular fermion pair.  

The analytic expressions for the $h\to f\bar{f}$ decay rates,  given as computer files with the arXiv submission 
of this work, will be useful in  future precision analysis of these decay modes in effective field theory, 
as well as for benchmarking all-purpose tools for automated NLO SMEFT calculations as they  become available.  

\section*{Acknowledgements}
The research of J.M.C.~is supported by an STFC Postgraduate Studentship. B.P.~is grateful to the Weizmann Institute of Science for its kind hospitality and support through the SRITP and the Benoziyo Endowment Fund for the Advancement of Science.

\appendix

\section{Decoupling constants}
\label{sec:DecouplingConstants}

The \msbar-renormalized input parameters in Eq.~(\ref{eq:InputPars}) are defined in a five-flavor version of QED$\times$QCD,
where particles with masses at the EW scale are integrated out.  These are related to the corresponding quantities
in SMEFT through decoupling constants.  Adapting the notation of \cite{Cullen:2019nnr} to the current paper, we 
define these decoupling constants as
\begin{align}
\label{eq:DecoupRelations}
\overline{m}^{\rm SM}_f(\mu)& = \zeta_f(\mu,m_t, m_H,M_W,M_Z) 
\overline{m}_f(\mu)  \, , \nonumber \\
\overline{e}^{\rm SM}(\mu) &= \zeta_e(\mu,m_t, m_H,M_W,M_Z) 
\overline{e}(\mu) \, ,
\end{align}
where the parameters with the superscript ``SM'' include all particles in the renormalization procedure. The decoupling
constants are calculated as a double series in the operator expansion and in perturbation theory.  In order to obtain 
the decay rate in terms of our chosen input parameters, the prescription is to first do the calculation in the full SM,
then include terms involving decoupling constants to convert to the \msbar-renormalized parameters in five-flavor
QED$\times$QCD. The exact relations are
\begin{align}
\label{eq:GammaDec}
\Gamma_f^{(4,0)} &= \Gamma_{f,\rm{SM}}^{(4,0)} \, , \nonumber \\
\Gamma_f^{(6,0)} &= \Gamma_{f,\rm{SM}}^{(6,0)} \, , \nonumber \\
\Gamma_f^{(4,1)} &= \Gamma_{f,\rm{SM}}^{(4,1)} +
2 \Gamma_{f}^{(4,0)}\left(\zeta_f^{(4,1)}+\zeta_e^{(4,1)}\right)
 \, , \nonumber \\
 \Gamma_f^{(6,1)} &=  \Gamma_{f,\rm{SM}}^{(6,1)} +
 2\Gamma_f^{(4,0)}\left(\zeta_f^{(6,1)}+\zeta_e^{(6,1)}\right)
 +2\Gamma_f^{(6,0)} \zeta_f^{(4,1)} 
 \nonumber \\*
 &+ \sqrt{2} C_{fH}\frac{\overline{v}^3}{\overline{m}_f} \Gamma_f^{(4,0)}  \left( \zeta_f^{(4,1)} +\zeta_e^{(4,1)} \right) \,,
\end{align}
 where in all cases the expressions on the right-hand side are evaluated using the parameters in Eq.~(\ref{eq:InputPars})
 directly.

The decoupling constants for the electric charge and the $b$-quark mass were given in \cite{Cullen:2019nnr}.  Here we give results for the decoupling constants for the $\tau$-lepton 
and $c$-quark masses.   In writing the results, it is convenient to split up tadpole and non-tadpole contributions.  This separation is gauge dependent, although the sum of the terms and the decoupling
constants themselves are gauge independent.  We define the tadpole contributions in Feynman gauge, making use 
of the quantities
\begin{align}
T_\text{ Feyn.}^{(4)} &= \frac{1}{32 \pi^2 \overline{v}} \bigg\{ \hat{A}_0(M_W^2)(12 M_W^2 +2m_H^2)- 8M_W^4 + \hat{A}_0(M_Z^2) (6 M_Z^2 +m_H^2)  \nonumber \\
	&-4 M_Z^4 + 3m_H^2 \hat{A}_0(m_H^2)-8 N_c m_t^2 \hat{A}_0(m_t^2) \bigg\} \, , \nonumber \\
T_\text{Feyn.}^{(6)}  &= \frac{\overline{v}}{32 \pi^2} \bigg\{4 \left(C_{H\Box}-\frac{C_{HD}}{4}\right)m_H^2\left( \hat{A}_0(m_H^2)-\hat{A}_0(M_W^2) \right) - 2 C_{H \Box} m_H^2 \hat{A}_0(M_Z^2)  \nonumber \\
	&-6 C_H \overline{v}^2 \hat{A}_0 (m_H^2) + \left(24 \hat{A}_0(M_W^2)-16 M_W^2 \right) C_{HW} M_W^2 + \left( 3 \hat{A}_0(M_Z^2) - 2M_Z^2 \right) M_Z^2 \nonumber \\
	&\times \left[ C_{HD}+4 \left( C_{HW} \hat{c}_w^2 +C_{HB}\hat{s}_w^2 + C_{HWB} \hat{c}_w \hat{s}_w \right) \right] + 4 N_c  \sqrt{2} \overline{v} m_t C_{tH} \hat{A}_0(m_t^2) \bigg\} \nonumber \\
	&+\overline{v}^2  \left[ C_{H\Box}-\frac{C_{HD}}{4}+ \frac{\hat{c}_w}{\hat{s}_w} \left( C_{HWB} + \frac{\hat{c}_w}{4 \hat{s}_w} C_{HD} \right) \right]T_\text{Feyn.}^{(4)} \, ,
\end{align}
where the superscript represents a contribution at mass dimension $i$ and 
\begin{align}
\hat{A}_0(m^2) = m^2 + m^2 \log \left(\frac{\mu^2}{m^2} \right) \, .
\end{align}
For the $\tau$-lepton and charm-quark masses in the SM, the decoupling constants are
\begin{align}
\zeta_{m_\tau}^{(4,1)} &=\frac{1}{64 \pi^2 \overline{v}^2} \left\{58 M_W^2-40 M_W^2 \hat{c}_w^2-21 M_Z^2+4 M_W^2 \ln\left( \frac{\mu^2}{M_W^2} \right) + \left(72 M_W^2 \nonumber \right. \right. \\
	 &- \left. \left. 48 M_W^2 \hat{c}_w^2 -22 M_Z^2 \right) \ln\left( \frac{\mu^2}{M_Z^2} \right) \right\} - \frac{1}{\overline{v} m_H^2}T_{\text{Feyn.}}^{(4)} \, ,
\end{align}

\begin{align}
\zeta_{m_c}^{(4,1)} &=\frac{1}{576 \pi^2 \overline{v}^2} \left\{182 M_W^2-160 M_W^2 \hat{c}_w^2-49 M_Z^2+36 M_W^2 \ln\left( \frac{\mu^2}{M_W^2} \right) + 6\left(40 M_W^2 \nonumber \right. \right. \\
	 &- \left. \left. 32 M_W^2 \hat{c}_w^2 -5 M_Z^2 \right) \ln\left( \frac{\mu^2}{M_Z^2} \right) \right\} - \frac{1}{\overline{v} m_H^2}T_{\text{Feyn.}}^{(4)}  \, .
\end{align}
The dimension-6 contributions to the $m_f$ decoupling constants can be written in the generic form
\begin{align}
\label{eq:SMEFTDecouplingNoTadBreakdown}
\zeta_{m_f}^{(6,1)}= - \frac{1}{m_H^2 \overline{v}} T^{(6)}_{\text{Feyn.}}  - \frac{1}{2 m_H^2 \overline{v} } \frac{\Gamma_f^{(6,0)}}{\Gamma_f^{(4,0)}} T^{(4)}_{\text{Feyn.}} + \zeta^{(6,1)}_{m_f,\text{ no-tad.}} \, ,
\end{align}

\begin{align}
\zeta_{m_f, \, \text{no-tad}}^{(6,1)} &= \zeta_{m_f, \, \text{NL}}^{(6,1)}+\zeta_{m_f, \, \text{LH}}^{(6,1)} \log \left( \frac{\mu^2}{m_H^2} \right)+\zeta_{m_f, \, \text{LW}}^{(6,1)} \log \left( \frac{\mu^2}{M_W^2} \right)+\zeta_{m_f, \, \text{LZ}}^{(6,1)} \log \left( \frac{\mu^2}{M_Z^2} \right) \nonumber \\  
	&+\zeta_{m_f, \, \text{Lt}}^{(6,1)} \log \left( \frac{\mu^2}{m_t^2} \right) \, .
\end{align}
For $f=\tau$ we find
\begin{align}
\zeta^{(6,1)}_{m_\tau,\text{NL}} &= - \frac{1}{128 \pi^2  M_W M_Z^3 \hat{s}_w^3} \left\{-3 C_{HD}  M_W M_Z \hat{s}_w\left(6 M_W^4+7M_Z^4 -14 M_W^2 M_Z^2 \right) \nonumber \right. \\
	&+ 12 C_{HWB} M_W^2 M_Z^2 \hat{s}_w^2 \left(4 M_W^2-3 M_Z^2 \right)+2 M_Z^2 \hat{s}_w^3 \left[ \frac{6 \sqrt{2}}{\hat{s}_w} \frac{\overline{v}\overline{e}}{\overline{m}_\tau} C_{\tau W}  M_W M_Z \nonumber \right. \\
	 &\times \left(2 M_W^2-M_Z^2 \right) +2 \sqrt{2} \frac{\overline{v}  \bar{e}}{\overline{m}_\tau} C_{\tau B} M_Z^2 \left(4 M_W^2-3 M_Z^2 \right) + 8 N_c \frac{m_t^3}{\overline{m}_\tau} C^{(1)}_{\substack{lequ \\ 3333}} M_W M_Z  \nonumber \\
	 &+ \sqrt{2} \frac{\overline{v}}{\overline{m}_\tau} C_{\tau H}  M_W M_Z \left(3 m_H^2 +2 M_W^2 +M_Z^2 \right)+2 M_W M_Z \left( \left( 10 M_W^2-7 M_Z^2 \right) C^{(1)}_{\substack{Hl \\ 33}} \right. \nonumber \\
	 &+ \left. \left. \left. \left( 16 M_W^2-7 M_Z^2 \right) C^{(3)}_{\substack{Hl \\ 33}}+\left( 10 M_W^2-8 M_Z^2 \right) C_{H\tau} \right) \right] \right\} \, , \\
\zeta_{m_\tau, \, \text{LH}}^{(6,1)} &= -\frac{3 \sqrt{2} m_H^2 }{64 \pi^2 } \frac{\overline{v}}{\overline{m}_\tau} C_{\tau H} \, , \\
\zeta_{m_\tau, \, \text{LW}}^{(6,1)} &= \frac{\hat{c}_w^2}{32 \pi^2  M_Z \hat{s}_w^3} \left\{ 4 C_{HWB} M_W M_Z^2 \hat{s}_w^2+C_{HD} M_W^2 M_Z \hat{s}_w \nonumber \right. \\
	&- \left. \sqrt{2} \frac{\overline{v}}{\overline{m}_\tau} M_Z^3 \hat{s}_w^2  \left(6 C_{\tau W} \bar{e}+ C_{\tau H} \hat{s}_w \right) \right\} \, , \\
\zeta_{m_\tau, \, \text{LZ}}^{(6,1)} &= - \frac{1}{64 \pi^2 M_W M_Z^3 \hat{s}_w^3} \left\{ C_{HD}  M_W M_Z \hat{s}_w \left(22 M_W^2 M_Z^2 -12 M_W^4 - 11 M_Z^4 \right) \nonumber \right. \\
	&+   4 C_{HWB}  M_W^2 M_Z^2 \hat{s}_w^2 \left( 6 M_W^2-7 M_Z^2 \right) + \sqrt{2} \frac{\overline{v}}{\overline{m}_\tau} C_{\tau H}  M_W M_Z^5 \hat{s}_w^3  \nonumber  \\ 
	&+ 6 \sqrt{2} \frac{\overline{v}\overline{e}}{\overline{m}_\tau} M_Z^4 \hat{s}_w^2 \left( C_{\tau B} \hat{s}_w + C_{\tau W} \hat{c}_w \right)  \left(4 M_W^2-3 M_Z^2 \right) \nonumber \\ 
	&- \left. 24 \left(C^{(1)}_{\substack{Hl \\ 33}}+C^{(3)}_{\substack{Hl \\ 33}} \right)  M_W M_Z^5 \hat{s}_w^5 + 12 C_{H \tau}  M_W M_Z^3 \hat{s}_w^3 \left( 2 M_W^2-M_Z^2 \right) \right\} \, , \\
\zeta_{m_\tau, \, \text{Lt}}^{(6,1)} &= - \frac{N_c}{8 \pi^2 } \frac{m_t^3}{\overline{m}_\tau} C^{(1)}_{\substack{lequ \\ 3333}} \, .
\end{align}
For $f=c$ one finds
\begin{align}
\zeta^{(6,1)}_{m_c,\text{NL}} &= - \frac{1}{1152 \pi^2 M_Z^3 \hat{s}_w^3} \left\{ C_{HD} M_Z \hat{s}_w \left(98 M_W^2 M_Z^2-49 M_Z^4-22M_W^4 \right) \nonumber \right. \\
	&-  12 C_{HWB} M_W \left(17 M_Z^4 + 26 M_W^4 -43 M_W^2 M_Z^2 \right) + 18 \sqrt{2} \frac{\overline{v}}{\overline{m}_c} C_{cH}  M_Z^3 \hat{s}_w^3 \nonumber \\
	&\times \left(3 m_H^2 +2 M_W^2 +M_Z^2 \right) + 12 M_Z^3 \hat{s}_w^3 \left[- \frac{\sqrt{2}}{\hat{c}_w} \frac{\overline{e} \overline{v}}{\overline{m}_c} C_{cB} \left(8 M_W^2-5 M_Z^2 \right) \nonumber \right. \\
	&+ \frac{\sqrt{2}}{\hat{s}_w} \frac{\overline{e} \overline{v}}{\overline{m}_c} C_{cW} \left(14 M_W^2-5 M_Z^2 \right)- C^{(1)}_{\substack{Hq \\ 22}}  \left( 20 M_W^2 -11 M_Z^2 \right) \nonumber \\
	&+C^{(3)}_{\substack{Hq \\ 22}} \left( 38 M_W^2 -11 M_Z^2 \right) - 2 C_{Hc} \left(10 M_W^2- 7 M_Z^2 \right) \nonumber \\
	&+ \left. \left. 12 \frac{m_t^3}{\overline{m}_c} \left( C^{(1)}_{\substack{qu \\ 2332}}+ C_{F} C^{(8)}_{\substack{qu \\ 2332}}  \right) \right] \right\} \, , \\
\zeta_{m_c, \, \text{LH}}^{(6,1)} &= -\frac{3 \sqrt{2} m_H^2}{64  \pi^2} \frac{\overline{v}}{\overline{m}_c} C_{c H} \, , \\
\zeta_{m_c, \, \text{LW}}^{(6,1)} &= \frac{1}{32 \pi^2 M_Z^5 \hat{s}_w^3} \left\{ C_{HD}  M_W^4 M_Z^3 \hat{s}_w +4 C_{HWB} M_W^3 M_Z^4 \hat{s}_w^2 \right. \nonumber \\
	 &- \left. \sqrt{2} \frac{\overline{v}}{\overline{m}_c} C_{cH}  M_W^2 M_Z^5 \hat{s}_w^3 - 6 \sqrt{2} \frac{\overline{e} \overline{v}}{\overline{m}_c} C_{cW} M_W^2 M_Z^5 \hat{s}_w^2  \right\} \, , \\
\zeta_{m_c, \, \text{LZ}}^{(6,1)} &= \frac{1}{192 \pi^2 M_W M_Z^5 \hat{s}_w^3} \left\{ C_{HD} M_W M_Z^3 \hat{s}_w \left(8 M_W^4 +5 M_Z^4 -10 M_W^2 M_Z^2 \right) \right. \nonumber \\
	&+ 12 C_{HWB} M_W^2 M_Z^4 \hat{s}_w^2 \left(5 M_Z^2 - 4 M_W^2 \right) - 3 \sqrt{2} \frac{\overline{v}}{\overline{m}_c} C_{cH} M_W M_Z^7 \hat{s}_w^3 \nonumber \\
	&+ 6 \sqrt{2} \frac{\overline{e} \overline{v}}{\overline{m}_c}  M_Z^6 \hat{s}_W^2 \left(8 M_W^2-5 M_Z^2  \right) \left( C_{cB}\hat{s}_w- C_{cW} \hat{c}_w \right) \nonumber \\
	&+ \left. 48 M_W M_Z^7 \hat{s}_w^5 \left(C^{(3)}_{\substack{Hq \\ 22}} - C^{(1)}_{\substack{Hq \\ 22}} \right) + 12 C_{Hc} M_W M_Z^5 \hat{s}_w^3 \left( 4 M_W^2-M_Z^2 \right) \right\} \, , \\
\zeta_{m_c, \, \text{Lt}}^{(6,1)} &= -\frac{1}{4 \pi^2} \frac{m_t^3}{m_c} \left\{C^{(1)}_{\substack{qu \\ 2332}} + C_F  C^{(8)}_{\substack{qu \\ 2332}}\right\} \, .
\end{align}

\newpage
\begin{table}
\begin{center}
\small
\begin{minipage}[t]{4.45cm}
\renewcommand{\arraystretch}{1.5}
\begin{tabular}[t]{c|c}
\multicolumn{2}{c}{$1:X^3$} \\
\hline
$Q_G$                & $f^{ABC} G_\mu^{A\nu} G_\nu^{B\rho} G_\rho^{C\mu} $ \\
$Q_{\widetilde G}$          & $f^{ABC} \widetilde G_\mu^{A\nu} G_\nu^{B\rho} G_\rho^{C\mu} $ \\
$Q_W$                & $\epsilon^{IJK} W_\mu^{I\nu} W_\nu^{J\rho} W_\rho^{K\mu}$ \\ 
$Q_{\widetilde W}$          & $\epsilon^{IJK} \widetilde W_\mu^{I\nu} W_\nu^{J\rho} W_\rho^{K\mu}$ \\
\end{tabular}
\end{minipage}
\begin{minipage}[t]{2.7cm}
\renewcommand{\arraystretch}{1.5}
\begin{tabular}[t]{c|c}
\multicolumn{2}{c}{$2:H^6$} \\
\hline
$Q_H$       & $(H^\dag H)^3$ 
\end{tabular}
\end{minipage}
\begin{minipage}[t]{5.1cm}
\renewcommand{\arraystretch}{1.5}
\begin{tabular}[t]{c|c}
\multicolumn{2}{c}{$3:H^4 D^2$} \\
\hline
$Q_{H\Box}$ & $(H^\dag H)\Box(H^\dag H)$ \\
$Q_{H D}$   & $\ \left(H^\dag D_\mu H\right)^* \left(H^\dag D_\mu H\right)$ 
\end{tabular}
\end{minipage}
\begin{minipage}[t]{2.7cm}

\renewcommand{\arraystretch}{1.5}
\begin{tabular}[t]{c|c}
\multicolumn{2}{c}{$5: \psi^2H^3 + \hbox{h.c.}$} \\
\hline
$Q_{eH}$           & $(H^\dag H)(\bar l_p e_r H)$ \\
$Q_{uH}$          & $(H^\dag H)(\bar q_p u_r \widetilde H )$ \\
$Q_{dH}$           & $(H^\dag H)(\bar q_p d_r H)$\\
\end{tabular}
\end{minipage}

\vspace{0.25cm}

\begin{minipage}[t]{4.7cm}
\renewcommand{\arraystretch}{1.5}
\begin{tabular}[t]{c|c}
\multicolumn{2}{c}{$4:X^2H^2$} \\
\hline
$Q_{H G}$     & $H^\dag H\, G^A_{\mu\nu} G^{A\mu\nu}$ \\
$Q_{H\widetilde G}$         & $H^\dag H\, \widetilde G^A_{\mu\nu} G^{A\mu\nu}$ \\
$Q_{H W}$     & $H^\dag H\, W^I_{\mu\nu} W^{I\mu\nu}$ \\
$Q_{H\widetilde W}$         & $H^\dag H\, \widetilde W^I_{\mu\nu} W^{I\mu\nu}$ \\
$Q_{H B}$     & $ H^\dag H\, B_{\mu\nu} B^{\mu\nu}$ \\
$Q_{H\widetilde B}$         & $H^\dag H\, \widetilde B_{\mu\nu} B^{\mu\nu}$ \\
$Q_{H WB}$     & $ H^\dag \sigma^I H\, W^I_{\mu\nu} B^{\mu\nu}$ \\
$Q_{H\widetilde W B}$         & $H^\dag \sigma^I H\, \widetilde W^I_{\mu\nu} B^{\mu\nu}$ 
\end{tabular}
\end{minipage}
\begin{minipage}[t]{5.2cm}
\renewcommand{\arraystretch}{1.5}
\begin{tabular}[t]{c|c}
\multicolumn{2}{c}{$6:\psi^2 XH+\hbox{h.c.}$} \\
\hline
$Q_{eW}$      & $(\bar l_p \sigma^{\mu\nu} e_r) \sigma^I H W_{\mu\nu}^I$ \\
$Q_{eB}$        & $(\bar l_p \sigma^{\mu\nu} e_r) H B_{\mu\nu}$ \\
$Q_{uG}$        & $(\bar q_p \sigma^{\mu\nu} T^A u_r) \widetilde H \, G_{\mu\nu}^A$ \\
$Q_{uW}$        & $(\bar q_p \sigma^{\mu\nu} u_r) \sigma^I \widetilde H \, W_{\mu\nu}^I$ \\
$Q_{uB}$        & $(\bar q_p \sigma^{\mu\nu} u_r) \widetilde H \, B_{\mu\nu}$ \\
$Q_{dG}$        & $(\bar q_p \sigma^{\mu\nu} T^A d_r) H\, G_{\mu\nu}^A$ \\
$Q_{dW}$         & $(\bar q_p \sigma^{\mu\nu} d_r) \sigma^I H\, W_{\mu\nu}^I$ \\
$Q_{dB}$        & $(\bar q_p \sigma^{\mu\nu} d_r) H\, B_{\mu\nu}$ 
\end{tabular}
\end{minipage}
\begin{minipage}[t]{5.4cm}
\renewcommand{\arraystretch}{1.5}
\begin{tabular}[t]{c|c}
\multicolumn{2}{c}{$7:\psi^2H^2 D$} \\
\hline
$Q_{H l}^{(1)}$      & $(H^\dag i\overleftrightarrow{D}_\mu H)(\bar l_p \gamma^\mu l_r)$\\
$Q_{H l}^{(3)}$      & $(H^\dag i\overleftrightarrow{D}^I_\mu H)(\bar l_p \sigma^I \gamma^\mu l_r)$\\
$Q_{H e}$            & $(H^\dag i\overleftrightarrow{D}_\mu H)(\bar e_p \gamma^\mu e_r)$\\
$Q_{H q}^{(1)}$      & $(H^\dag i\overleftrightarrow{D}_\mu H)(\bar q_p \gamma^\mu q_r)$\\
$Q_{H q}^{(3)}$      & $(H^\dag i\overleftrightarrow{D}^I_\mu H)(\bar q_p \sigma^I \gamma^\mu q_r)$\\
$Q_{H u}$            & $(H^\dag i\overleftrightarrow{D}_\mu H)(\bar u_p \gamma^\mu u_r)$\\
$Q_{H d}$            & $(H^\dag i\overleftrightarrow{D}_\mu H)(\bar d_p \gamma^\mu d_r)$\\
$Q_{H u d}$ + h.c.   & $i(\widetilde H ^\dag D_\mu H)(\bar u_p \gamma^\mu d_r)$\\
\end{tabular}
\end{minipage}

\vspace{0.25cm}

\begin{minipage}[t]{4.75cm}
\renewcommand{\arraystretch}{1.5}
\begin{tabular}[t]{c|c}
\multicolumn{2}{c}{$8:(\bar LL)(\bar LL)$} \\
\hline
$Q_{ll}$        & $(\bar l_p \gamma_\mu l_r)(\bar l_s \gamma^\mu l_t)$ \\
$Q_{qq}^{(1)}$  & $(\bar q_p \gamma_\mu q_r)(\bar q_s \gamma^\mu q_t)$ \\
$Q_{qq}^{(3)}$  & $(\bar q_p \gamma_\mu \sigma^I q_r)(\bar q_s \gamma^\mu \sigma^I q_t)$ \\
$Q_{lq}^{(1)}$                & $(\bar l_p \gamma_\mu l_r)(\bar q_s \gamma^\mu q_t)$ \\
$Q_{lq}^{(3)}$                & $(\bar l_p \gamma_\mu \sigma^I l_r)(\bar q_s \gamma^\mu \sigma^I q_t)$ 
\end{tabular}
\end{minipage}
\begin{minipage}[t]{5.25cm}
\renewcommand{\arraystretch}{1.5}
\begin{tabular}[t]{c|c}
\multicolumn{2}{c}{$8:(\bar RR)(\bar RR)$} \\
\hline
$Q_{ee}$               & $(\bar e_p \gamma_\mu e_r)(\bar e_s \gamma^\mu e_t)$ \\
$Q_{uu}$        & $(\bar u_p \gamma_\mu u_r)(\bar u_s \gamma^\mu u_t)$ \\
$Q_{dd}$        & $(\bar d_p \gamma_\mu d_r)(\bar d_s \gamma^\mu d_t)$ \\
$Q_{eu}$                      & $(\bar e_p \gamma_\mu e_r)(\bar u_s \gamma^\mu u_t)$ \\
$Q_{ed}$                      & $(\bar e_p \gamma_\mu e_r)(\bar d_s\gamma^\mu d_t)$ \\
$Q_{ud}^{(1)}$                & $(\bar u_p \gamma_\mu u_r)(\bar d_s \gamma^\mu d_t)$ \\
$Q_{ud}^{(8)}$                & $(\bar u_p \gamma_\mu T^A u_r)(\bar d_s \gamma^\mu T^A d_t)$ \\
\end{tabular}
\end{minipage}
\begin{minipage}[t]{4.75cm}
\renewcommand{\arraystretch}{1.5}
\begin{tabular}[t]{c|c}
\multicolumn{2}{c}{$8:(\bar LL)(\bar RR)$} \\
\hline
$Q_{le}$               & $(\bar l_p \gamma_\mu l_r)(\bar e_s \gamma^\mu e_t)$ \\
$Q_{lu}$               & $(\bar l_p \gamma_\mu l_r)(\bar u_s \gamma^\mu u_t)$ \\
$Q_{ld}$               & $(\bar l_p \gamma_\mu l_r)(\bar d_s \gamma^\mu d_t)$ \\
$Q_{qe}$               & $(\bar q_p \gamma_\mu q_r)(\bar e_s \gamma^\mu e_t)$ \\
$Q_{qu}^{(1)}$         & $(\bar q_p \gamma_\mu q_r)(\bar u_s \gamma^\mu u_t)$ \\ 
$Q_{qu}^{(8)}$         & $(\bar q_p \gamma_\mu T^A q_r)(\bar u_s \gamma^\mu T^A u_t)$ \\ 
$Q_{qd}^{(1)}$ & $(\bar q_p \gamma_\mu q_r)(\bar d_s \gamma^\mu d_t)$ \\
$Q_{qd}^{(8)}$ & $(\bar q_p \gamma_\mu T^A q_r)(\bar d_s \gamma^\mu T^A d_t)$\\
\end{tabular}
\end{minipage}

\vspace{0.25cm}

\begin{minipage}[t]{3.75cm}
\renewcommand{\arraystretch}{1.5}
\begin{tabular}[t]{c|c}
\multicolumn{2}{c}{$8:(\bar LR)(\bar RL)+\hbox{h.c.}$} \\
\hline
$Q_{ledq}$ & $(\bar l_p^j e_r)(\bar d_s q_{tj})$ 
\end{tabular}
\end{minipage}
\begin{minipage}[t]{5.5cm}
\renewcommand{\arraystretch}{1.5}
\begin{tabular}[t]{c|c}
\multicolumn{2}{c}{$8:(\bar LR)(\bar L R)+\hbox{h.c.}$} \\
\hline
$Q_{quqd}^{(1)}$ & $(\bar q_p^j u_r) \epsilon_{jk} (\bar q_s^k d_t)$ \\
$Q_{quqd}^{(8)}$ & $(\bar q_p^j T^A u_r) \epsilon_{jk} (\bar q_s^k T^A d_t)$ \\
$Q_{lequ}^{(1)}$ & $(\bar l_p^j e_r) \epsilon_{jk} (\bar q_s^k u_t)$ \\
$Q_{lequ}^{(3)}$ & $(\bar l_p^j \sigma_{\mu\nu} e_r) \epsilon_{jk} (\bar q_s^k \sigma^{\mu\nu} u_t)$
\end{tabular}
\end{minipage}
\end{center}
\caption{\label{op59}
The 59 independent baryon number conserving dimension-6 operators built from Standard Model fields, in 
the notation of \cite{Jenkins:2013zja}.  The subscripts $p,r,s,t$ are flavor indices, and $\sigma^I$ are Pauli
matrices.}
\end{table}

\providecommand{\href}[2]{#2}\begingroup\raggedright\endgroup

\end{document}